\documentclass[%
 reprint,
 amsmath,amssymb,
 aps,
prstab,
]{revtex4-2}

\usepackage{graphicx}
\usepackage{dcolumn}
\usepackage{bm}
\usepackage[greek, english]{babel}
\usepackage{upgreek}

\begin{document}

\title{Design Study of a Dielectric Laser Undulator}

\author{Steffen A. Schmid}
 \email{steffen.schmid@tu-darmstadt.de}
\author{Uwe Niedermayer}
 \email{niedermayer@temf.tu-darmstadt.de}

\affiliation{Technical University Darmstadt,\\ Institute for Accelerator Science and Electromagnetic Fields,\\ Schlossgartenstrasse 8, 64289 Darmstadt, Germany}

\date{\today}

\begin{abstract}
    Dielectric laser acceleration (DLA) achieves remarkable gradients from the optical near fields of a grating structure. Tilting the dielectric grating with respect to the electron beam leads to deflection forces and the DLA structure can be utilized as a microchip undulator. We investigate the beam dynamics in such structures analytically and by numerical simulations. A crucial challenge is to keep the beam focused, especially in direction of the narrow channel. An alternating phase focusing scheme is optimized for this purpose and matched lattice functions are obtained. We distinguish synchronous operation with phase jumps in the grating and asynchronous operation with a strictly periodic grating and well-designed synchronicity mismatch. Especially the asynchronous DLA undulator is a promising approach, since a simple, commercially available grating suffices for the focusing lattice design. We pave the way towards experiments of radiation generation in these structures and provide estimates of the emitted radiation wavelength and power. The analytical models are validated by numerical simulations in the dedicated DLA simulation tool \emph{DLAtrack6D} and \emph{Astra}, where the underlying laser fields are computed by \emph{CST Studio}.
\begin{description}
\item[Keywords] Dielectric Laser Acceleration, Electromagnetic Undulator, Table Top Light Source
\end{description}
\end{abstract}

\maketitle

\section{Introduction}

    The working principle of modern accelerator light sources relies on undulators transforming the energy of an electron beam into short-wavelength photon radiation~\cite{sas_ref.Schmuser.2014}. Large-scale facilities like the E-XFEL~\cite{sas_ref.Altarelli.2006} and the Swiss-FEL~\cite{sas_ref.Milne.2017} use periodically alternating magnets with approximately $1\,\rm{T}$ field strength to induce a transverse wiggling motion of the particle beam. With a typical period length of about one centimeter, magnetic undulators reach several tens to hundreds of meters total length. Active undulator designs use microwave~\cite{sas_ref.Shintake.1983}, terahertz~\cite{sas_ref.Rohrbach.2019}, or optical~\cite{sas_ref.Toufexis.2015} electromagnetic fields to drive a wiggling motion of the electron beam. In order to miniaturize accelerator light sources, especially optical laser undulators are a promising concept since they potentially reach $\rm{GeV/m}$ acceleration and deflection gradients which allow to realize ultrashort undulator periods~\cite{sas_ref.Plettner.2008}.
    
    Dielectric laser acceleration (DLA) utilizes the periodic laser field distribution in a dielectric grating to accelerate electrons in excess of $1\,\rm{GV/m}$ field strength~\cite{sas_ref.England.2014}. Tilting the grating with respect to the electron beam direction induces transverse deflection forces such that the structure effectively serves as a DLA undulator~\cite{sas_ref.Plettner.2008b}. Thus, DLA technology has the potential to reduce the dimensions of an X-ray free electron laser to table-top size~\cite{sas_ref.Rosenzweig.2013}. However, owing to the tiny dimensions and the laser induced defocusing forces beam transport through a DLA lattice remains challenging~\cite{sas_ref.Niedermayer.2017}. Magnetic quadrupole or solenoid lenses can neither mitigate the inherent beam divergence within the aperture nor compensate the defocusing induced by the nonlinear laser fields. Thus, DLA undulators require an entirely laser-based focusing scheme as described in Refs.~\cite{sas_ref.Naranjo.2012,sas_ref.Niedermayer.2018,sas_ref.Niedermayer.2020}. This article discusses how such a focusing scheme, i.e. alternating phase focusing (APF), can be adapted to DLA undulator structures, where the reference particle moves on an undulating trajectory. The ensemble of all other particles is then supposed to execute stable betatron oscillations around the reference particle. To achieve this, the APF for DLA formalism~\cite{sas_ref.Niedermayer.2018,sas_ref.Niedermayer.2020} is generalized to periodically curved reference trajectories. Moreover, borrowing from spatial harmonic focusing~\cite{sas_ref.Naranjo.2012}, mismatch in the synchronicity condition leads to a drifting synchronous phase which is modeled as an asynchronous APF scheme.
    
    This design study is intended to pave the way from the conceptual idea~\cite{sas_ref.Plettner.2008} towards a first experiment generating radiation from DLA undulators. Therefore, the study adopts machine parameters of the accelerator research experiment (ARES) at the R\&D facility for short innovative bunches and accelerators at DESY (SINBAD)~\cite{sas_ref.Dorda.2016,sas_ref.Marchetti.2020}. ARES generates a $E = 107\,\rm{MeV}$ electron beam with $Q= 0.5\,\rm{pC}$ bunch charge, $\sigma_{\rm{t}} = 0.75\,\rm{fs}$ bunch length, $\varepsilon_{\rm{x/y}} = 1\,\rm{nm}$ transverse geometric emittance, and $\sigma_{\rm{E}} / E = 0.06\,\%$ energy spread. Furthermore, the silica grating design assumes $\lambda_0 = 2\,\rm{\upmu m}$  drive-laser wavelength as commonly used for DLA experiments in the accelerator on a chip (ACHIP) collaboration~\cite{sas_ref.Cankaya.2021,sas_ref.Shiloh.2021,sas_ref.Hermann.2019,sas_ref.Leedle.2018}.
    
    The structure of the paper is as follows. Section~\ref{sec:tilted_grating} recapitulates the theory of tilted DLA structures from Refs.~\cite{sas_ref.Plettner.2009, sas_ref.Niedermayer.2017} and provides an optimized grating design for a high-gradient fused silica DLA undulator. Section~\ref{sec:undulator_lattice} discusses the single particle dynamics for both a synchronous and an asynchronous DLA undulator lattice. The investigation provides analytical formulas for the DLA undulator parameter $K_{\rm{und}}$. Section~\ref{sec:alternative_phase_focusing} generalizes the DLA alternating phase focusing scheme~\cite{sas_ref.Niedermayer.2018} to DLA undulators which are shown to provide a FODO lattice with the same periodicity as the undulator period. A subsequent beam-matching study reveals the specifications and limitations of the investigated DLA undulator designs. Finally, section~\ref{sec:particle_tracking} shows particle tracking simulations for the DLA design parameters that were identified as most suitable with respect to experimental prospects. The paper concludes with an outlook on the feasibility of a DLA undulator experiment at ARES and its potential implications on future applications of DLA undulators in large-scale facilities, such as the E-XFEL.

\section{\label{sec:tilted_grating}Tilted DLA Grating Cell}

\begin{figure}
        \includegraphics{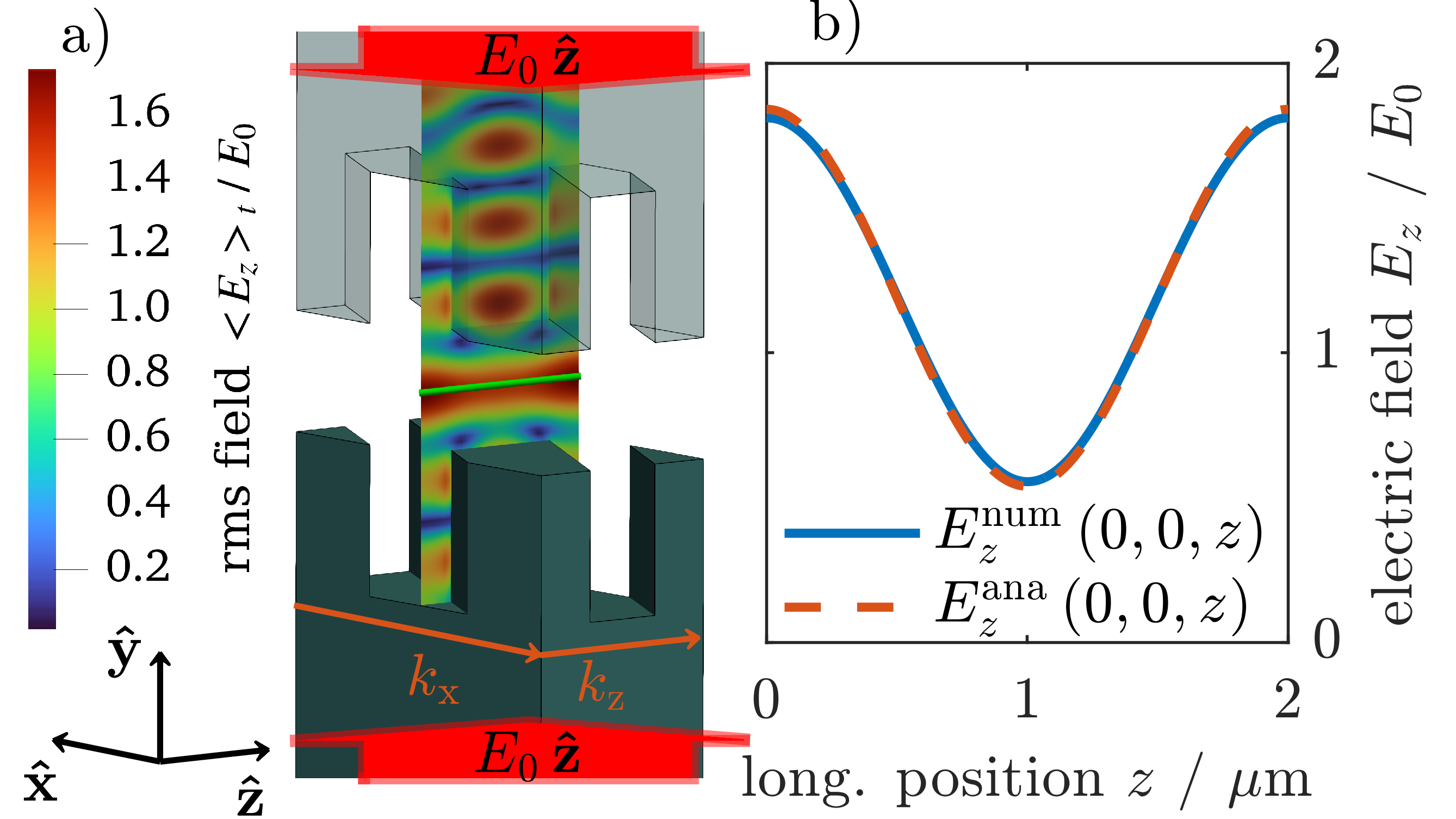}
        \caption{\label{figure1_schmid_prab} \emph{CST Studio} simulation~\cite{sas_ref.3DS.2021} of the field between two opposing dielectric gratings for a $\bm{\hat{z}}$ polarized drive-laser with amplitude $E_0$ in a). In b), the field strength is compared to a cosine wave with $e_1(0,0)$ complex amplitude, showing that the fundamental Fourier coefficient is dominant.}
    \end{figure}

    Laser-driven DLA undulators utilize Lorentz forces induced by optical near fields in tilted dielectric gratings to generate a wiggling beam trajectory~\cite{sas_ref.Plettner.2008b}. Figure~\ref{figure1_schmid_prab}\,a) visualizes the laser field in one tilted DLA grating cell simulated using \emph{CST Studio}~\cite{sas_ref.3DS.2021}. The working principle of DLA undulators relies either on synchronous or on asynchronous grating structures. In a synchronous structure the beam particles travel along the green line exactly one grating cell per laser period in $\hat{\bm{z}}$ direction. Thus, the beam velocity $v_{\rm{z}} = \beta c$, the longitudinal grating constant $\lambda_{\rm{z}} = 2\pi/k_{\rm{z}}$, and the laser wavelength $\lambda_{\rm{0}} = 2\pi/k_{\rm{0}}$ fulfill the synchronicity condition
    \begin{equation}
     \lambda_{\rm{z}} = \beta \, \lambda_{\rm{0}}~\rm{.} \label{eqn:phase_synchronous_condition}
    \end{equation}
    In an asynchronous structure the laser wavelength $\lambda_{\rm{0}}$ does not match \eqref{eqn:phase_synchronous_condition} which introduces a smooth phase drift of the optical fields at the particle positions. The following investigation provides an overview of the basic properties for both synchronous and asynchronous DLA undulators.
    
     The reciprocal grating vector $\bm{k}_{\rm{g}} = k_{\rm{x}} \,\hat{\bm{x}} + k_{\rm{z}} \,\hat{\bm{z}}$ in Fig.~\ref{figure1_schmid_prab}\,a) is parallel to the $y=0$ symmetry plane and tilted by an angle $\alpha$ with respect to the beam path along $\hat{\bm{z}}$. The periodic modulation of the laser field $E_{\rm{z}}^{\rm{num}}$ along $\bm{\hat{z}}$ in Fig.~\ref{figure1_schmid_prab}\,b) mainly corresponds to the fundamental harmonic of the reciprocal grating vector $k_{\rm{z}}$. Thus, the  Fourier coefficient
    \begin{equation}
        e_1(x,y) = \lambda_{\rm{z}}^{-1}\,\int_{-\lambda_{\rm{z}}/2}^{\lambda_{\rm{z}}/2} \tilde{E}_{\rm{z}} \left(x,y,z\right) e^{i k_{\rm{z}}z} dz \label{eqn:fourier_coefficient_e1}
    \end{equation}
    with the phasor $\tilde{E}_{\rm{z}}$ for the drive-laser field characterizes the effective interaction field $E_{\rm{z}}^{\rm{ana}}$ of the DLA undulator. 

    A general beam dynamics description based on the non-linear, time-dependent Lorentz force in such setups is rather cumbersome. However, in phase-synchronous structures the Panofsky-Wenzel theorem~\cite{sas_ref.Panofsky.1956} allows to simplify the analysis. Referring to the investigation in Ref.~\cite{sas_ref.Niedermayer.2017}, the effective electromagnetic interaction corresponds to a scalar potential
    \begin{multline}
        V\left(x,y,z,t;\varphi\right) = -\frac{q \left|e_1(0,0)\right|}{k_{\rm{z}}} \cosh\left(k_{\rm{y}} y\right)\\
        \sin\left[k_{\rm{x}}\, x + k_{\rm{z}} \left(z - \beta \,ct\right) + \varphi\right]\label{eqn::DLA_potential_synchronous}
    \end{multline}
    with
    \begin{align}
        k_{\rm{x}} &= k_{\rm{z}} \tan{\alpha}\label{eqn:gratingvector_kx}~\rm{,}\\
        k_{\rm{y}} &\equiv \sqrt{\left|{k_{0}}^2 - {k_{\rm{x}}}^2 - {k_{\rm{z}}}^2\right|}\label{eqn:gratingvector_ky}~\rm{,}
    \end{align}
    the particle charge $q$, and an arbitrary phase $\varphi$. Unlike the definition of $V$ used in Ref.~\cite{sas_ref.Niedermayer.2017}, the force $-\nabla V$ acting on a synchronous particle with $z = s + \beta\, ct$ and $y=0$ is constant in time and anti-/parallel to the grating vector $\bm{k}_{\rm{g}}$. Hence, electrons passing a tilted diffraction grating experience both transverse deflection and longitudinal acceleration. The relative phase $\varphi$ between the electron beam and the laser field determines the sign and strength of the force.
    
\begin{figure}
        \includegraphics{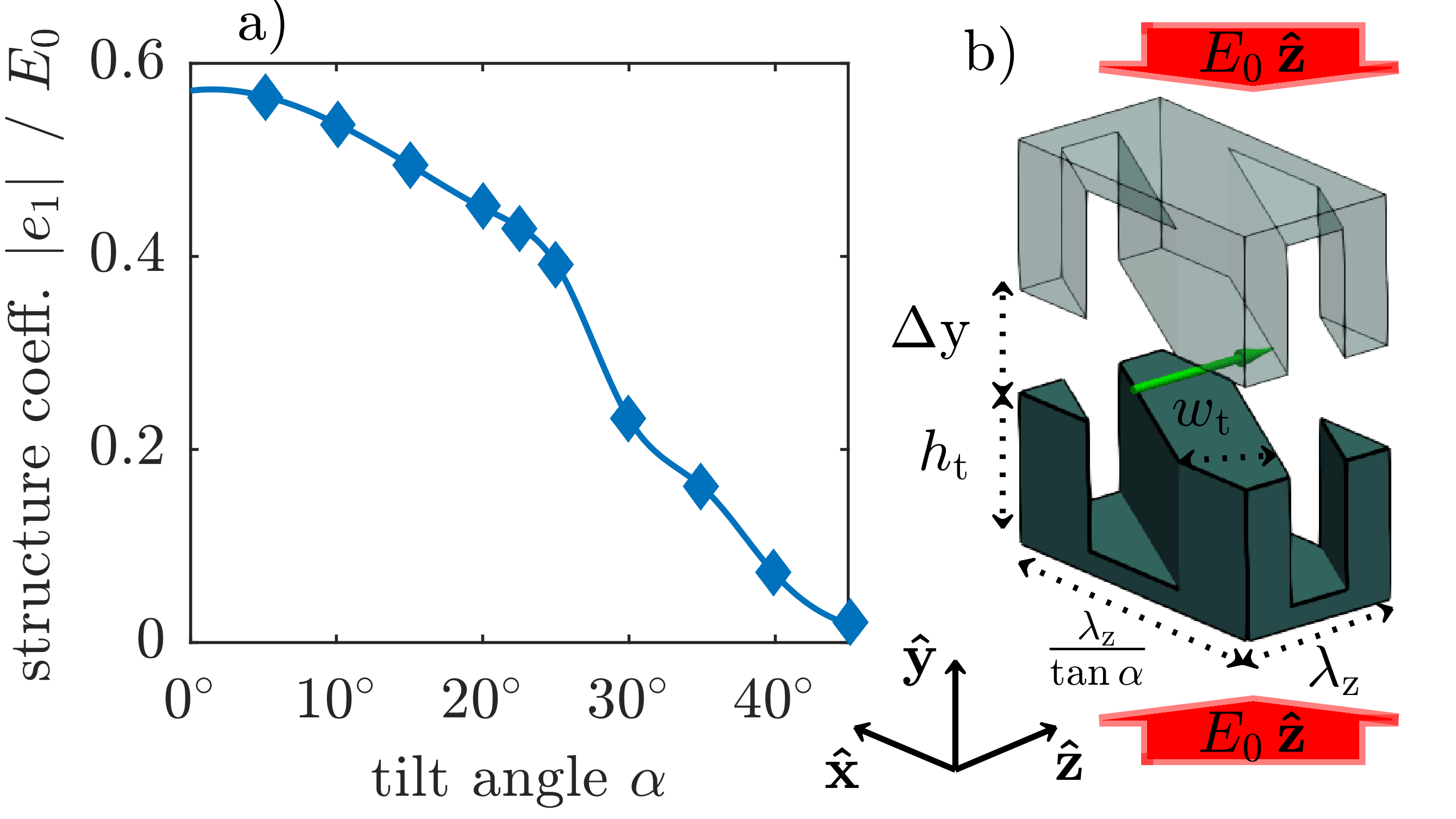}
        \caption{\label{figure2_schmid_prab} \emph{CST Studio} simulation~\cite{sas_ref.3DS.2021} for the structure coefficient $\left|e_{\rm{1}}\left(\alpha\right)\right|$ of a silica grating at different tilt angles $\alpha$ for a $\bm{\hat{z}}$ polarized drive-laser with amplitude $E_0$ in a). The grating design in b) uses a tooth width of $w_{\rm{t}} = 0.6 \lambda_{\rm{g}}$, a tooth height of $h_{\rm{t}} = 0.75 \lambda_0$, and a gap width of $\Delta y = 0.6 \lambda_0$}.
    \end{figure}

    Effective electromagnetic interaction between the drive-laser field and the electrons requires maximizing the structure coefficient $\left|e_{\rm{1}}\left(\alpha\right)\right|$ defined as \eqref{eqn:fourier_coefficient_e1} evaluated at $x = y = 0$ for the grating tilt angle $\alpha$. Figure~\ref{figure2_schmid_prab}\,a) shows the results of a numerical optimization study for the geometry in Fig.~\ref{figure2_schmid_prab}\,b) using \emph{CST Studio}~\cite{sas_ref.3DS.2021}. The DLA cell consist of two opposing silica diffraction gratings with $\epsilon_{\rm{r}} = 2.0681$ for a laser wavelength of $\lambda_0 = 2\,\rm{\upmu m}$ and a longitudinal grating wave-number $k_{\rm{z}}$ according to \eqref{eqn:phase_synchronous_condition} for $\beta \approx 1$. The transverse grating wave number $k_{\rm{x}}$ gets larger as the tilt angle $\alpha$ increases. The optima for the grating tooth width $w_{\rm{t}} = 1.0\,\rm{\upmu m}$ and for the tooth height $h_{\rm{t}} = 1.5\,\rm{\upmu m}$ are similar to typical specifications of commercially available NIR spectrometer gratings~\cite{sas_ref.IbsenPhotonics.2020}. In Fig.~\ref{figure2_schmid_prab}\,a) the coefficient $\left|e_{\rm{1}}\left(\alpha\right)\right|$ is maximum at zero tilt and decreases as $\alpha$ increases. Qualitatively, this degradation is a consequence of the laser field attenuation from the inner grating surface towards the beam path indicated as a green arrow. Combining \eqref{eqn:phase_synchronous_condition}, \eqref{eqn:gratingvector_kx}, and \eqref{eqn:gratingvector_ky} yields the field attenuation constant $k_{\rm{y}} = k_0 \sqrt{\beta^{-2}\cos^{-2}{\alpha} - 1}$ in $\hat{\bm{y}}$ direction. The attenuation limits the maximum beam channel width to $\Delta y \sim \lambda_0$ for which the numerical geometry optimization yields a good compromise at $\Delta y = 1.2\,\rm{\upmu m}$.

\section{\label{sec:undulator_lattice}DLA Undulator Lattice}

     In order to achieve a wiggling beam trajectory, the deflection needs to switch sign twice per undulator period length~\cite{sas_ref.Plettner.2008}. This can be implemented on a tilted grating structure in two ways: either, one can introduce phase jumps, as in an alternating phase focusing scheme for a synchronous lattice~\cite{sas_ref.Niedermayer.2018}, or one can introduce a continuous phase drift by a mismatch of the laser wavelength and the grating period in \eqref{eqn:phase_synchronous_condition} resulting in an asynchronous lattice.
     
\begin{figure}
        \includegraphics{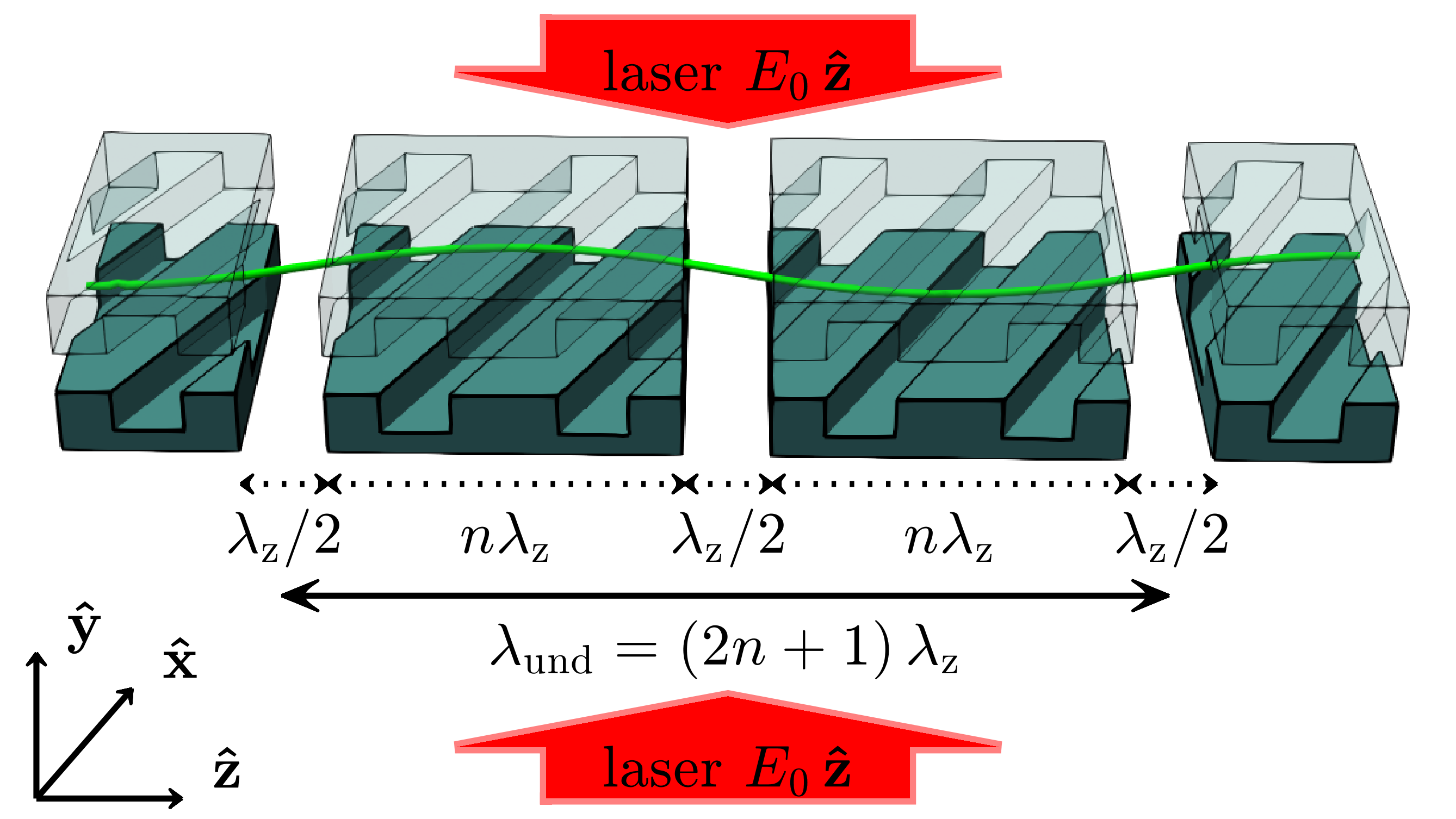}
        \caption{\label{figure3_schmid_prab}Lattice design and beam trajectory (amplitude not to scale) of a phase-synchronous DLA undulator for a $\bm{\hat{z}}$ polarized drive-laser with amplitude $E_0$. Each segment consisting of $n = 2$ tilted DLA cells imposes a constant deflection force on the beam. A subsequent drift space $\lambda_{\rm{z}}/2$ shifts the phase by $\Delta \varphi = \pi$ which flips the direction of the force.}
    \end{figure}

     Figure~\ref{figure3_schmid_prab} shows a lattice for the synchronous scheme with $\Delta z = \lambda_{\rm{z}}/2$ drift spaces equivalent to phase shifts $\Delta \varphi = \pi$ in order to accomplish the alternating deflection. Hence, a full undulator period comprises two blocks of $n \in \mathbb{N}^+$ DLA grating cells separated by $\lambda_{\rm{z}}/2$ which corresponds to the undulator wavelength $\lambda_{\rm{und}} = \left(2 n + 1\right) \lambda_{\rm{z}}$. As it is the case for magnetic undulators, the DLA undulator requires $\lambda_{\rm{und}}/4$ sections at its entrance and exit to avoid imposing a transverse drift motion on the beam~\cite{sas_ref.Wille.2009}.
    
\begin{figure}
        \includegraphics{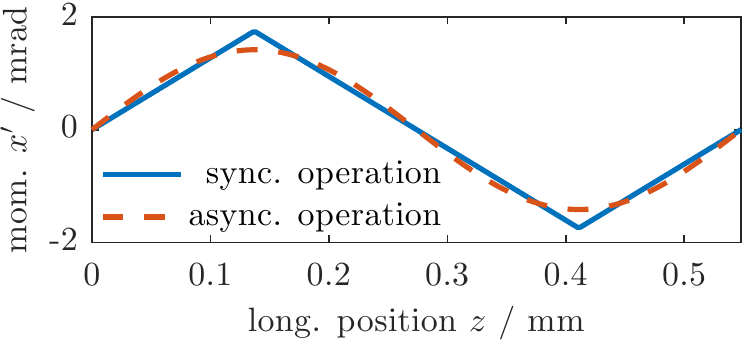}
        \caption{\label{figure4_schmid_prab} Transverse momentum modulation of the reference particle in a tilted grating undulator. In a synchronous DLA lattice, the trajectory is a sequence of constant acceleration with alternating sign. The asynchronous DLA lattice imposes a sinusoidal varying deflection.}
    \end{figure}

    The blue line in Fig.~\ref{figure4_schmid_prab} visualizes the transverse deflection $x^{\prime} = dx/dz$ across one wavelength $\lambda_{\rm{und}} = 554\,\rm{\upmu m}$ for the reference particle located at $\left(x_0,y_0 = 0,s_0;\varphi_0\right)$ in a synchronous DLA undulator with $n = 138$ cells per segment. The trajectory corresponds to a periodic sequence of $\lambda_{\rm{und}}/2$ sections in which the particle experiences the acceleration $dp_{\rm{x}}/dt = \pm \partial V/\partial x$. Since the longitudinal velocity is almost constant, $z\left(t\right) \approx s_0 + \beta \, ct$, and the transverse deflection remains negligibly small, $\left|x\left(t\right) - x_0\right| \ll 2\pi/k_x$, the motion resembles the dynamics of a uniformly accelerated particle. If applicable, direct integration of $p_{\rm{x}} = \int \partial V/\partial x \,dt$ with $\partial V/\partial x$ evaluated at $\left(x_0,0,z\left(t\right);\varphi_0\right)$ yields an analytical expression for $x^{\prime}\left(z\right)$ which is the piece-wise linear function in Fig.~\ref{figure4_schmid_prab}.
    
    Referring to the particle motion in conventional magnetic undulators~\cite{sas_ref.Wille.2009} with $x^{\prime} = K_{\rm{und}}/\gamma \, \sin\left(2\pi z/\lambda_{\rm{und}}\right)$, the equivalent undulator parameter $K_{\rm{und}}$ for the synchronous DLA lattice may be obtained from the fundamental Fourier coefficient of $x^{\prime}\left(z\right)$ in Fig.~\ref{figure4_schmid_prab} as
    \begin{align}
        K_{\rm{und}}^{\rm{sync}} &= \frac{2}{\lambda_{\rm{und}}} \int_0^{\lambda_{\rm{und}}} \gamma x^{\prime}\left(z\right) \sin\left(\frac{2\pi}{\lambda_{\rm{und}}} z\right)dz \nonumber \\
        &= \frac{2}{\pi^2} \frac{q\,\left|e_1\left(\alpha\right)\right|\tan{\alpha}}{m c^2} \frac{ \lambda_{\rm{und}}}{\beta^2} \cos{\varphi_{\rm{s}}} \label{eqn:undulator_parameter_sync}
    \end{align}
    with the electron charge $q$, the rest mass $m$, and the synchronous phase 
    \begin{equation}\label{eqn:synchronous_phase_referenceparticle}
        \varphi_{\rm{s}} \equiv k_{\rm{x}}\,x_0 + k_{\rm{z}}\,s_0 + \varphi_0~\rm{.}
    \end{equation}
    Note that in contrast to Ref.~\cite{sas_ref.Niedermayer.2018}, where the synchronous phase is a lattice property, the synchronous phase is determined by the injection properties of the highly stiff beam. Furthermore, the particle trajectory contains contributions of higher harmonics (HH) with odd mode numbers $m_{\rm{HH}} \in \left\lbrace3,5,\dots\right\rbrace$ and amplitudes proportional to $1 / m_{\rm{HH}}^{2}$.
    
    Similar to its magnetic equivalent, the DLA undulator parameter is proportional to the undulator wavelength $\lambda_{\rm{und}}$ and the effective electromagnetic field strength $\left|e_{\rm{1}}\left(\alpha\right)\right|$. In addition, $K_{\rm{und}}^{\rm{sync}}$ varies with the grating tilt angle $\alpha$ and the synchronous phase $\varphi_s$ such that it vanishes completely if $\alpha = 0$ or $\varphi_s \in \left\lbrace90^{\circ},270^{\circ}\right\rbrace$. For the investigated silica structure it exists an optimum at $\alpha \approx 27^{\circ}$ which maximizes the undulator parameter $K_{\rm{und}}$.
    
    The orange dashed line in Fig.~\ref{figure4_schmid_prab} for $x'(z)$ of an asynchronous DLA undulator is sinusoidal and thus has only the fundamental Fourier component. Because of the inherent phase drift, the grating structure does not require additional drift spaces to flip the deflection force. The synchronous phase increases every grating cell length $\lambda_{\rm{z}}$ by $\Delta \varphi_s = 2\pi\, \lambda_{\rm{z}}/\lambda_{\rm{und}}$. Integration of the deflection force $dp_{\rm{x}}/dt=\partial V/ \partial x \propto \cos\left[2\pi/\lambda_{\rm{und}} \, z\left(t\right)\right]$ results in an undulator parameter 
    \begin{equation}
    K_{\rm{und}}^{\rm{async}} = \frac{1}{2 \pi} \frac{q\,\left|e_1\left(\alpha\right)\right|\tan{\alpha}}{m c^2} \frac{ \lambda_{\rm{und}}}{\beta^2}\label{eqn:undulator_parameter_async}\rm{.} \end{equation}
    In contrast to the synchronous undulator, $K_{\rm{und}}^{\rm{async}}$ represents a deflection which is approximately $20\,\%$ smaller than the maximum of $K_{\rm{und}}^{\rm{sync}}$ and does not depend on the synchronous phase $\varphi_{\rm{s}}$.
    
\begin{figure}
        \includegraphics{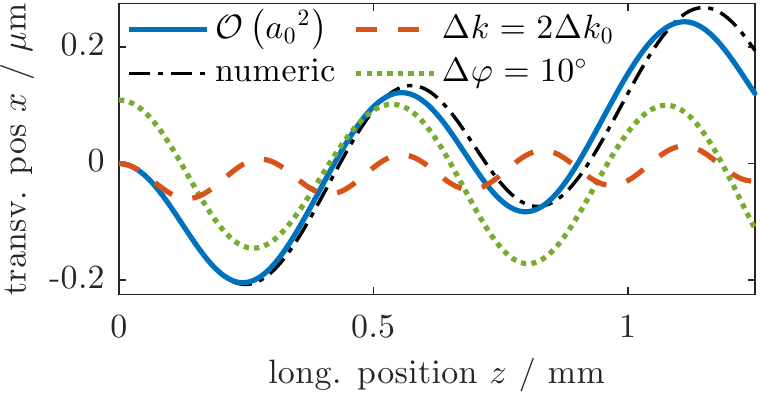}
        \caption{\label{figure5_schmid_prab} Numerical and $\mathcal{O}\left({a_0}^{-2}\right)$ analytically approximated solutions for the reference trajectories $x\left(z\right)$ in an asynchronous DLA undulator with $\Delta k_0 / k_0 = 0.36\%$ detuning from synchronous operation, $\alpha = 27^{\circ}$ tilt angle, and $E_0 = 10\,\rm{GV/m}$ laser field strength. Both the laser detuning $\Delta k$ and the initial phase $\Delta \varphi$ influence the dynamics.}
    \end{figure}

    Modeling the beam dynamics in asynchronous DLA structures requires additional caution since the aforementioned approximation of the electromagnetic interaction as a scalar potential \eqref{eqn::DLA_potential_synchronous} is not necessarily valid. A complete description of the relativistic electron dynamics in an asynchronous DLA laser field premises on the more general Hamiltonian~\cite{sas_ref.Kroll.1981}
    \begin{multline}
        H\left(x,p_x,y,p_y,ct,\gamma;z\right) = - p_{\rm{z}} = \\
        - \sqrt{\gamma^2 - 1 - \left(p_x - a_x\right)^2 - \left(p_y - a_y\right)^2} - a_z ~\rm{.}\label{eqn:hamilton_vetorpotential}
    \end{multline}
    with the canonical momenta $\bm{p}$ in units of $m c$ and the scaled vector potential $\bm{a} \equiv q/\left(mc\right) \bm{A}$. Referring to Appx.~\ref{apx:asynchronous_vectorfield_ansatz}, the geometry of an asynchronous DLA undulator motivates a vector potential in the form of
    \begin{equation}\label{eqn:representation_vetorpotential}
        \bm{a} = a_0 \begin{pmatrix} \xi_{\rm{x}} \cosh\left(k_{\rm{y}} y\right) \sin\left(k_{\rm{z}} z + k_{\rm{x}} x - k_0 ct + \varphi_0\right) \\ \xi_{\rm{y}} \sinh\left(k_{\rm{y}} y\right) \cos\left(k_{\rm{z}} z + k_{\rm{x}} x - k_0 ct + \varphi_0\right) \\ \xi_{\rm{z}} \cosh\left(k_{\rm{y}} y\right) \sin\left(k_{\rm{z}} z + k_{\rm{x}} x - k_0 ct + \varphi_0\right)\end{pmatrix}
    \end{equation}
    with a unitless amplitude
    \begin{equation}
        a_0 \equiv \frac{q E_0/k_0}{m c^2}\label{eqn:hamilton_vectorpotential_amplitude}
    \end{equation}
    proportional to the laser field strength $E_0$ and a polarization vector $\bm{\xi}$. The injection phase $\varphi_0$ is defined by the time at which the reference particle enters the undulator.
    As discussed in Appx.~\ref{apx:asynchronous_perturbation_theory}, the dynamics of \eqref{eqn:hamilton_vetorpotential} yields a coupled system of six nonlinear differential equations which can be solved numerically as shown in Fig.~\ref{figure5_schmid_prab}.
    
    Typically, the energy variation across one laser wavelength $q E_0 \lambda_0$ is much smaller than the electron rest energy $m c^2$. A laser field strength in range of the silica damage threshold $E_0 \sim 1-10\,\rm{GV/m}$~\cite{sas_ref.England.2014, sas_ref.Soong.2012} at $\lambda_0 = 2\,\rm{\upmu m}$ in \eqref{eqn:hamilton_vectorpotential_amplitude} yields $a_0 \sim 10^{-3}$. Following the idea of Ref.~\cite{sas_ref.Tran.1987}, the particle dynamics might be approximated treating the laser field as a perturbation in $a_0 \ll 1$. The approximation of \eqref{eqn:hamilton_vetorpotential} in Appx.~\ref{apx:asynchronous_perturbation_theory} for a relativistic electron beam including terms $\mathcal{O}\left({\gamma_0}^{-2}\right)$ and $\mathcal{O}\left({a_0}^2\right)$ yields for the transverse coordinate of the reference particle
    \begin{multline}
         x\left(z\right) \approx \left(\xi_{\rm{x}} a_0 + \xi_{\rm{z}} \frac{a_0 k_{\rm{x}}}{k_{\rm{und}}} \right) \frac{\left[\cos{\varphi} - \cos{\left(k_{\rm{und}} \, z - \varphi\right)}\right]}{\gamma_0 k_{\rm{und}}}\\
         + x_0 + \xi_{\rm{z}} \frac{a_0}{\gamma_0} \frac{k_{\rm{x}}}{k_{\rm{und}}} \sin{\varphi} \, z + \underbrace{\dots}_{\mathcal{O}\left({a_0}^2\right)}\label{eqn:hamilton_vectorpotenital_transversex}
    \end{multline}
    with $k_{\rm{und}} = 2 \pi/\lambda_{\rm{und}}$ as the undulator wave number
    \begin{equation}
        k_{\rm{und}} = \left(1 + \frac{1}{2{\gamma_0}^2}\right) k_0 - k_{\rm{z}} \approx \frac{k_0}{\beta_0} - k_{\rm{z}}\label{eqn:hamilton_vectorpotenital_undulatorwavelength}
    \end{equation}
    and the initial phase offset at $z_0=0$ defined as
    \begin{equation}\label{eqn:asynchronous_phase}
        \varphi = k_{\rm{x}}\, x_0 - k_0 \, ct_0 + \varphi_0 ~\rm{.}
    \end{equation}
    The approximation \eqref{eqn:hamilton_vectorpotenital_transversex} evaluated for the design parameters in Fig.~\ref{figure5_schmid_prab} shows a reasonable agreement with the numerically computed solution of \eqref{eqn:hamilton_vetorpotential}. The analytical perturbation approach also reproduces fundamental aspects of the beam dynamics and provides basic insights on the parameter dependencies in asynchronous DLA undulators as follows:

    First, the undulator wavelength $\lambda_{\rm{und}}$ in \eqref{eqn:hamilton_vectorpotenital_undulatorwavelength} corresponds to the beat wave due to the deviation from the synchronicity condition \eqref{eqn:phase_synchronous_condition}. Hence, variation of the drive-laser wavelength $\lambda_0$ in an experiment allows direct adjustment of $\lambda_{\rm{und}}$. The orange dashed line in Fig.~\ref{figure5_schmid_prab} shows the effect of doubling the laser detuning to $\Delta k_0 / k_0 = 0.72\%$ which reduces the effective undulator wavelength by a factor $1/2$.
    
    Second, the amplitude of the cosine terms in \eqref{eqn:hamilton_vectorpotenital_transversex} represents the asynchronous DLA undulator parameter $K_{\rm{und}}^{\rm{async}}$. The first term $\xi_{\rm{x}} a_0$ with $a_0 \propto \lambda_0$ represents the deflection induced by the transverse field component. Electromagnetic RF~\cite{sas_ref.Shintake.1983}, MIR~\cite{sas_ref.Toufexis.2015}, and THz~\cite{sas_ref.Rohrbach.2019} driven undulators use this effect to wiggle the particle beam. Since DLA structures use NIR wavelengths $\lambda_0 \sim 2\,\rm{\upmu m}$ and typically $k_{\rm{x}} \gg k_{\rm{und}}$ the influence of the first term is negligibly small compared to the second term $\xi_{\rm{z}} a_0 k_{\rm{x}}/k_{\rm{und}}$. Identifying $k_{\rm{x}} = \tan{\alpha}\,k_{\rm{z}}$ and $\xi_{\rm{z}} E_0 = \left|e_1\left(\alpha\right)\right|$ the second term reproduces the undulator parameter \eqref{eqn:undulator_parameter_async} in the relativistic limit $\beta \approx 1$. Hence, for the investigated DLA structures the simplified modeling approach based on the scalar potential $V$ provides reasonable results and, thus, can be a-posteriori justified. Further analysis of \eqref{eqn:hamilton_vetorpotential} in Ref.~\cite{sas_ref.Niedermayer.2022} indicates that our simulation code \emph{DLAtrack6D}~\cite{sas_ref.Niedermayer.2017} based on the scalar potential obtained by the Panofsky-Wenzel theorem~\cite{sas_ref.Panofsky.1956} provides an adequate tool to model the beam dynamics not only in synchronous, but also in asynchronous DLA structures generally. Using \emph{DLAtrack6D}, parameter studies for various DLA undulator designs can be performed in a numerically efficient way, giving rise to simple brute-force optimization.
    
    Third, equation \eqref{eqn:hamilton_vectorpotenital_transversex} shows that the initial phase offset $\varphi$ creates an initial deflection angle, i.e. a transverse drift proportional to $K_{\rm{und}}/\gamma\,\sin{\varphi}$. The green dotted line in Fig.~\ref{figure5_schmid_prab} shows how an offset $\Delta x = 0.1\,\rm{\upmu m}$ for $\alpha = 27^{\circ}$ tilt angle, equivalent to $\Delta \varphi = 10^{\circ}$, influences the transverse drift angle. Furthermore, a second order effect, $\mathcal{O}\left({a_0}^{-2}\right)$, results in the drift motion which is present in the analytical solution (blue line) as well as in the numerical solution (black line). Thus, for a finite bunch length, the asynchronous DLA undulator introduces a transverse beam divergence in $\hat{\bm{x}}$ direction which fundamentally limits the length of the undulator.

\section{\label{sec:alternative_phase_focusing}Alternating-Phase Focusing}
    
    Due to the small aperture in the $y$-coordinate, external beam optics elements cannot compensate the laser induced defocusing forces~\cite{sas_ref.Niedermayer.2018}. Consequently, an entirely laser-based focusing scheme such as APF~\cite{sas_ref.Niedermayer.2018,sas_ref.Niedermayer.2020} is required as otherwise the electron beam widens and most particles will be lost in the grating material. The following discussion generalizes the APF scheme for a sinusoidal wiggling reference trajectory.
    
    A paraxial approximation of the Hamiltonian in co-moving coordinates~\cite{sas_ref.Niedermayer.2018} yields
    \begin{equation}
        H\left(\bm{p}, \bm{\delta r};t\right) = \frac{1}{2 \gamma m} \left({p_{\rm{x}}}^2 + {p_{\rm{y}}}^2 + \frac{{p_{\rm{s}}}^2}{\gamma^2}\right) + V_{\rm{s}}\left(\bm{\delta r}\right)
    \end{equation}
    with ${\bm{\delta r}}^{\rm{T}} = \left(\delta x,\delta y,\delta s\right)$ representing the distance to a reference particle located at ${\bm{r}_0}^{\rm{T}} = \left(x_0,0,s_0 + \beta c t_0\right)$. The synchronous DLA potential
    \begin{multline}
        V_{\rm{s}}\left(\bm{\delta r}\right) \equiv V(\bm{\delta r} + \bm{r}_0,t_0;\varphi_0) + q \left|e_1\left(\alpha\right)\right| / {k_{\rm{z}}} \\
        \left[k_{\rm{z}} \left(\delta s + s0\right) + k_{\rm{x}} \left(\delta x + x_0\right)\right] \cos{\varphi_s}
    \end{multline}
    implies by definition a fixed point of motion at $\bm{\delta r} = 0$ where $-\nabla V_{\rm{s}} = 0$. In the vicinity of $\bm{r}_0$ including terms $\mathcal{O}\left(\bm{\delta r}^2\right)$ the synchronous DLA potential simplifies to
    \begin{multline}
         V_{\rm{s}}\left(\bm{\delta r}\right) \approx \frac{1}{2} q\,\left|e_1\left(\alpha\right)\right| \frac{k_0}{\beta} \sin{\varphi_s} \\
         \left[\left({\delta s} + \tan{\alpha} ~{\delta x}\right)^2 - \left(\gamma^{-2} +  {\tan{\alpha}}^2\right) {\delta y}^2\right]~\rm{.}
    \end{multline}
    which results in a coupled set of linear equations of motion
    \begin{gather}
        \frac{d^2}{dz^2} \bm{\delta r} = - \underbrace{\frac{k_0}{\gamma^3 \beta^3} \frac{q~\left|e_1\left(\alpha\right)\right|}{m c^2} \sin{\varphi_{\rm{s}}}}_{\equiv K_{\alpha}} ~ \hat{M}_{\alpha} \cdot \bm{\delta r} \label{eq:FODO:Hillsequation}\\
        \text{with~}\hat{M}_{\alpha} = \begin{pmatrix} \gamma^2 \tan^2{\alpha} & 0 & \gamma^2 \tan{\alpha} \\ 0 & -\left(1+\gamma^2 \tan^2{\alpha}\right) & 0 \\ \tan{\alpha} & 0 & 1 \end{pmatrix}\text{.}\nonumber
    \end{gather}
    
     In accordance with Ref.~\cite{sas_ref.Niedermayer.2018}, at zero tilt angle $\alpha = 0$ equation~\eqref{eq:FODO:Hillsequation} decouples to Hill's equations for a quadrupole lens~\cite{sas_ref.Reiser.2008} aligned in $y/s$ direction with focusing constant $K_{\alpha = 0}$. For a tilted grating with $\alpha~\neq~0$, diagonalization of $\hat{M}_{\alpha}$ provides analytical solutions for the beam dynamics. The tilted grating acts as a thick defocusing quadrupole lens in $y$-direction, which yields the general solution
     \begin{equation}
         {\delta y}\left(z\right) = c_1 \cosh{\sqrt{K} z} + c_2 \sinh{\sqrt{K} z},
     \end{equation}
     where tilt angle dependent focusing constant is given by 
     \begin{equation}
        K\left(\alpha\right) = K_{\alpha} \left(1 + \gamma^2 \tan^2{\alpha}\right) \label{eqn:focusing_strength_tilted_grating} \\
    \end{equation}
    and $c_i$ are constants to be determined from the initial conditions.
    
\begin{figure}
        \includegraphics{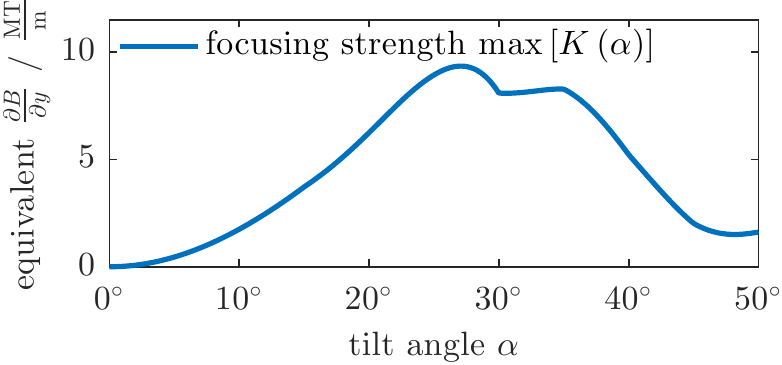}
        \caption{\label{figure6_schmid_prab} Tilt angle dependent quadrupole focusing of a synchronous DLA undulator at maximum $K\left(\alpha\right)$ for $\varphi_{\rm{s}} = \pi/2$ and $E_0 = 10\,\rm{GV/m}$ field amplitude with an optimum at $\alpha \approx 27^{\circ}$.}
    \end{figure}

    Figure~\ref{figure6_schmid_prab} shows a plot of the focusing constant of a synchronous DLA undulator, inserting the results of Fig.~\ref{figure1_schmid_prab}\,a) into \eqref{eqn:focusing_strength_tilted_grating}. 
    The maximum at $\alpha \approx 27^{\circ}$ tilt angle originates from a compromise between increasing $\tan{\alpha}$ and the thereby decreasing $\left|e_{\rm{1}}\left(\alpha\right)\right|$. The electron trajectory in $x$-/$s$-direction
    \begin{align}
        {\delta x}\left(z\right) &= -\left(c_3 + c_4 \, z\right) \cot{\alpha} \nonumber\\
        &\phantom{=} + \gamma^2\tan{\alpha} \left(c_5 \cos{\sqrt{K} z} + c_6 \sin{\sqrt{K} z}\right)\text{,}\\
        {\delta s}\left(z\right) &= c_3 + c_4 \, z + c_5 \cos{\sqrt{K} z} + c_6 \sin{\sqrt{K} z} ,
    \end{align} 
    is the superposition of a uniform drift motion and the particle dynamics in a focusing quadrupole lens. The drift direction ${\delta x}/{\delta s} = -\cot{\alpha}$ coincides with the translation symmetry axis of the grating which is perpendicular to $\bm{k}_{\rm{g}}$. Equivalent to a translation motion along the $x$-direction in a non-tilted grating, an electron drifting perpendicularly to $\bm{k}_{\rm{g}}$ in a tilted grating acquires zero phase advance ${\delta \varphi} = k_z\, {\delta s} + k_x\, {\delta x} = 0$ and its relative momentum $\bm{p}$ remains constant. Regarding the quadrupole-like terms, a longitudinal motion ${\delta s}$ translates directly to a transverse motion ${\delta x} = \gamma^2\tan{\alpha}\,{\delta s}$.

    The synchronous DLA undulator lattice shown in Fig.~\ref{figure3_schmid_prab} forms a sequence of focusing and defocusing sections effectively acting like a FODO channel. Each drift section of length $l_{\rm{O}} = \lambda_{\rm{z}}/2$ shifts the synchronous phase by $\Delta \varphi_{\rm{s}} = \pi$ which alternates sign of the focusing constant $\pm K\left(\alpha\right)$. Hence, the undulator wavelength $\lambda_{\rm{und}}$ determines the length of the de-/focusing sections as $l_{\rm{D/F}} = \lambda_{\rm{und}}/2 - l_{\rm{O}}$. An eigenvector analysis $\hat{T} \bm{\eta} = \bm{\eta}$ for the FODO transport matrices $\hat{T}$ provides the Courant-Snyder parameters $\bm{\eta}^{T} = \left(\hat{\beta},\hat{\alpha},\hat{\gamma}\right)$~\cite{sas_ref.Rosenzweig.2003} of the electron beam matched into the DLA undulator lattice. Optimizing $l_{\rm{D/F}}$ with respect to the beam size $\sigma = \sqrt{\varepsilon \hat{\beta}}$ as discussed in Ref.~\cite{sas_ref.Niedermayer.2018} yields the undulator wavelength $\lambda_{\rm{und}}$ for which the maximum of the betafunction $\hat{\beta}_{\rm{max}} = \rm{max}\left[\hat{\beta}\left(z\right)\right]$ is at its minimum value.
    
\begin{figure}
        \includegraphics{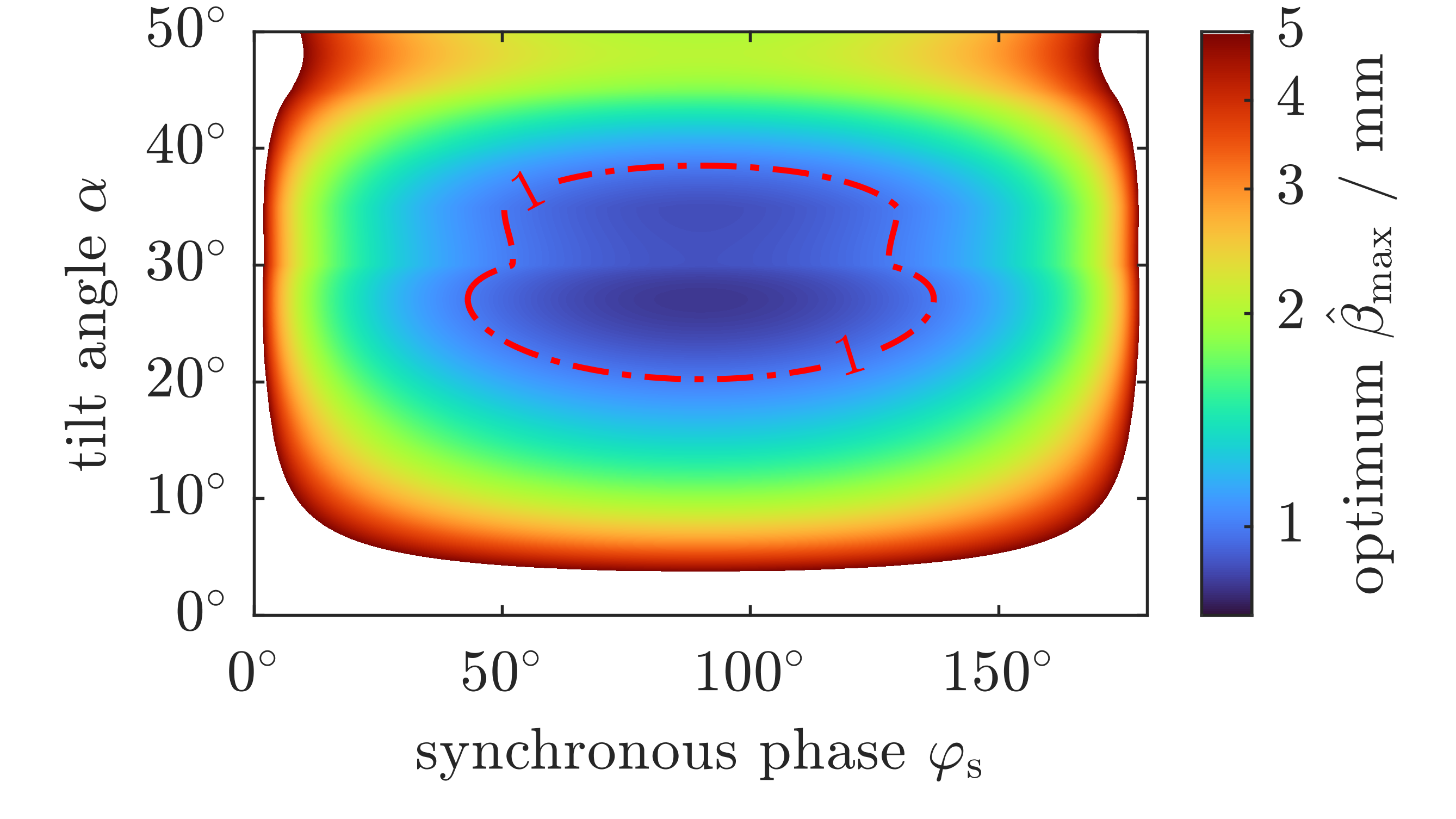}
        \caption{\label{figure7_schmid_prab}Maximum of the betafunction $\hat{\beta}_{\rm{max}}$ for a matched beam in a synchronous DLA undulator with $E_0 = 10\,\rm{GV/m}$ laser field strength at optimum FODO cell length. The absolute minimum of $\hat{\beta}_{\rm{max}}$ at $\varphi_{\rm{s}} = \pi/2$ implies $K_{\rm{und}}^{\rm{sync}}=0$.}
    \end{figure}

    Figure~\ref{figure7_schmid_prab} shows the minimum $\hat{\beta}_{\rm{max}}$ of a matched electron beam for $E_0 = 10\,\rm{GV/m}$ laser field strength. The smallest achievable value $\hat{\beta}_{\rm{max}} = 826\,\rm{\upmu m}$ appears for the maximum focusing constant at $\alpha = 27^{\circ}$ and $\varphi_{\rm{s}} = \pi/2$. However, according to \eqref{eqn:undulator_parameter_sync} the undulator parameter $K_{\rm{und}}^{\rm{sync}}$ vanishes at $\varphi_{\rm{s}} = \pi/2$. Conversely, the maximum undulator parameter $K_{\rm{und}}^{\rm{sync}}$ at $\varphi_{\rm{s}} \in \left\lbrace0,\pi\right\rbrace$ implies zero focusing $K\left(\alpha\right) = 0$ which is why $\hat{\beta}_{\rm{max}}$ diverges. Consequently, efficient operation of the synchronous DLA undulator requires a compromise between focusing and deflection strength, which is set via $\varphi_s$ by the injection properties (cf.~\eqref{eqn:synchronous_phase_referenceparticle}). For $\hat{\beta}_{\rm{max}} \approx 1\,\rm{mm}$ indicated with the red dashed line in Fig.~\ref{figure7_schmid_prab}, the optimal FODO lattice cell length yields $\lambda_{\rm{und}} \sim 0.6\,\rm{mm}$.
    
    An asynchronous DLA undulator effectively functions like a spatial harmonic focusing device as discussed in Ref.~\cite{sas_ref.Naranjo.2012}. The phase drift $\varphi_{\rm{s}} = k_{\rm{und}} \, z$ creates a continuously varying focusing function $K\left(\alpha, z\right) = K_0 \cos{\left(k_{\rm{und}} \, z\right)}$ with $K_0 \equiv \left.K\left(\alpha\right)\right|_{\varphi_{\rm{s}}=\pi/2}$. This results basically in the equation of motion for ponderomotive focusing in Ref.~\cite{sas_ref.Niedermayer.2022}. The efficiency of spatial harmonic focusing in DLA structures typically suffers form sharing the available laser power between the phase synchronous acceleration field and the phase asynchronous focusing field~~\cite{sas_ref.Naranjo.2012}. However, for the application in an undulator this drawback vanishes since the asynchronous field provides both beam wiggling and focusing. The total transport matrix $\hat{T}$ of an asynchronous DLA undulator is basically a matrix product of multiple single DLA cell transport matrices with $\Delta \varphi_{\rm{s}} = 2 \pi \lambda_{\rm{z}} / \lambda_{\rm{und}}$ phase shifts in between. Thus, a numerical evaluation of $\hat{T}$ facilitates the same eigenvector analysis for $\bm{\eta}$ as used for the synchronous design.
    
\begin{figure}
        \includegraphics{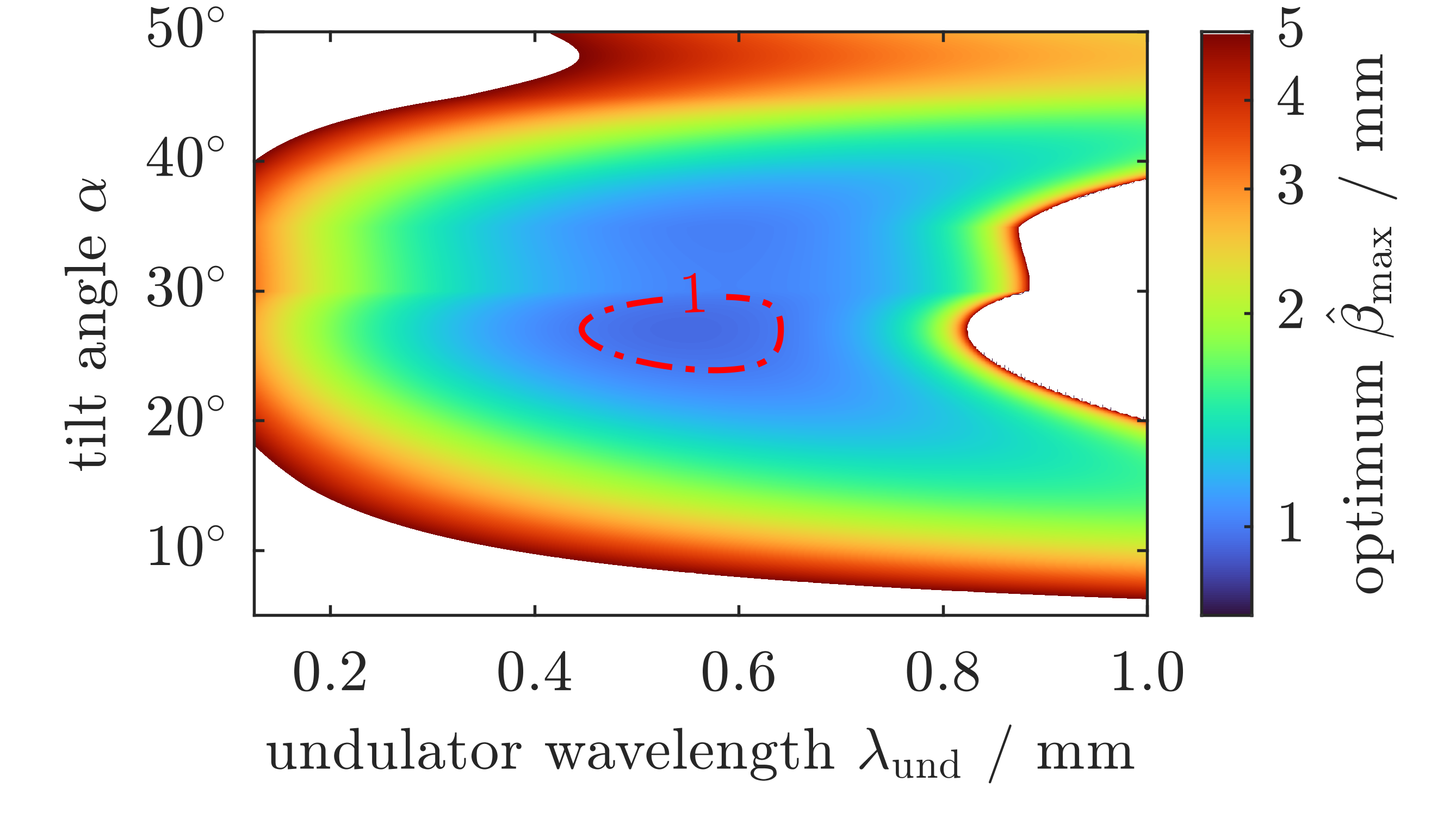}
        \caption{\label{figure8_schmid_prab} Maximum $\hat{\beta}_{\rm{max}}$ for a matched beam in an asynchronous DLA undulator with $E_0 = 10\,\rm{GV/m}$ laser field strength at optimum FODO cell length. In comparison to the synchronous design the minimum of $\hat{\beta}_{\rm{max}}$ at $\lambda_{\rm{und}} \approx 0.55\,\rm{mm}$ is larger, but the undulator parameter $K_{\rm{und}}^{\rm{async}}$ does not vanish.}
    \end{figure}

    Figure~\ref{figure8_schmid_prab} shows the minimum of $\hat{\beta}_{\rm{max}}$ for the asynchronous DLA undulator. The absolute minimum of $\hat{\beta}_{\rm{max}}$ is $952\,\rm{\upmu m}$ and appears for $\alpha = 27^{\circ}$ at $\lambda_{\rm{und}} \approx 0.55\,\rm{mm}$. In contrast to the synchronous design, the asynchronous undulator parameter $K_{\rm{und}}^{\rm{async}} \approx 0.3$ is nonzero for all injection phases. As in Ref.~\cite{sas_ref.Niedermayer.2018}, the betafunction diverges when $\lambda_{\rm{und}} \rightarrow 0$ and it also diverges for an undulator wavelength $\lambda_{\rm{und}} \gtrsim 0.8\,\rm{mm}$ when the FODO lattice reaches its over-focusing stability limit. Hence, the practical working range of the asynchronous DLA undulator indicated by the red dashed line restricts the achievable undulator wavelength to $0.45\,\rm{mm} \leq \lambda_{\rm{und}} \leq 0.64\,\rm{mm}$.
    
\begin{figure}
        \includegraphics{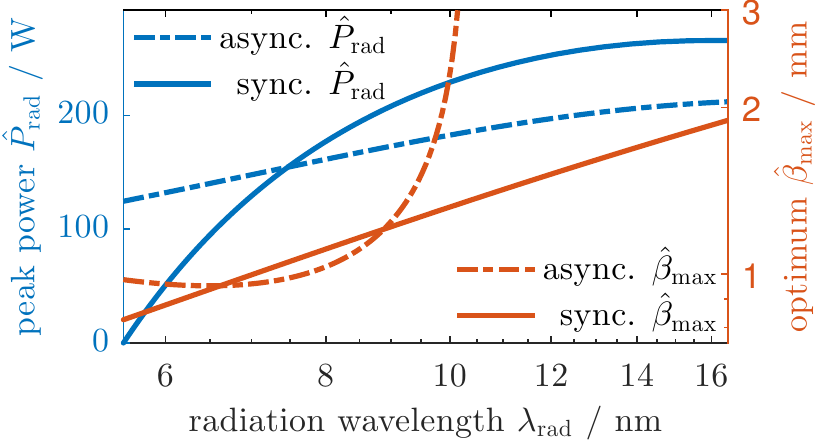}
        \caption{\label{figure9_schmid_prab} Comparison of the optimum $\hat{\beta}_{\rm{max}}$ for $\alpha = 27^{\circ}$, the radiation wavelength, and the radiation peak power for both the synchronous and the asynchronous DLA undulator design at $E_0 = 10\,\rm{GV/m}$ drive-laser field amplitude.}
    \end{figure}

    In order to compare the two lattice designs Fig.~\ref{figure9_schmid_prab} shows the optimum $\hat{\beta}_{\rm{max}}$ for $\alpha = 27^{\circ}$ tilt angle at the corresponding radiation wavelength~\cite{sas_ref.Wille.2009}
    \begin{equation}
        \lambda_{\rm{rad}} = \frac{\lambda_{\rm{und}}}{2 \gamma^2} \left( 1 + \frac{{K_{\rm{und}}}^2}{2} \right)
    \end{equation}
    and the maximum peak power~\cite{sas_ref.Attwood.1999}
    \begin{equation}
        \hat{P}_{\rm{rad}} \approx \frac{q \pi}{\varepsilon_0} \frac{\gamma^2 I}{\lambda_{\rm{und}}} \frac{{K_{\rm{und}}}^2}{1+{K_{\rm{und}}}^2}
    \end{equation}
    for the average current $I$ of a $Q = 0.5\,\rm{p C}$ electron bunch with $\sigma_{\rm{t}} = 0.75\,\rm{fs}$ pulse duration~\cite{sas_ref.Marchetti.2020}. Varying $\lambda_{\rm{und}}$ for the asynchronous DLA undulator directly affects the FODO lengths $l_{\rm{D/F}}$. However, the amplitude of the focusing function $K = K_0$ remains fixed. Hence, the optimum $\hat{\beta}_{\rm{max}}\left(\lambda_{\rm{rad}}\right)$ approaches a local minimum at $\lambda_{\rm{rad}} \approx 6.5\,\rm{nm}$ and starts diverging outside the interval $2\,\rm{nm} \lesssim \lambda_{\rm{rad}} \lesssim 10\,\rm{nm}$. Varying $\lambda_{\rm{und}}$ for the synchronous DLA undulator requires adjustment of the synchronous phase $\varphi_{\rm{s}}$ in a way that $\sqrt{K} \propto \sqrt{\sin{\varphi_{\rm{s}}}}\propto 1/\lambda_{\rm{und}}$ yields a constant betatron phase advance $\theta \equiv \sqrt{K} \, l_{\rm{D/F}}$, which facilitates to keep the betafunction $\hat{\beta}_{\rm{max}} \propto \frac{1}{\sqrt{K}}$ at its minimum, until $K$ reaches its upper limit $\rm{max}\left[K\left(\alpha\right)\right]$. Thus, a smaller radiation wavelength $\lambda_{\rm{rad}}$ yields smaller $\hat{\beta}_{\rm{max}}$. For equal peak radiation power, the optimum $\hat{\beta}_{\rm{max}}$ of the asynchronous design is slightly smaller than for the synchronous undulator. However, the synchronous design allows to trade off peak radiation power for a smaller value of $\hat{\beta}_{\rm{max}}$. 
    
    Table~\ref{tab:design_undulator_experiment} shows the specifications and performance parameters of both optimized DLA undulator designs for the ARES accelerator injection parameters. The synchronous DLA undulator yields the same peak power as the asynchronous design if both undulator parameters \eqref{eqn:undulator_parameter_sync} and \eqref{eqn:undulator_parameter_async} are equal which is the case for $\varphi_{\rm{s}} \approx 38^{\circ}$ synchronous phase. The damage threshold fluence $1.85\,\rm{J/cm^{2}}$ for a silica grating~\cite{sas_ref.K.Soong.2011} limits the maximum field strength to $E_0 \sim 10\,\rm{GV/m}$ and necessitates to utilize ultra-short laser pulses with $\tau \sim 100\,\rm{fs}$ pulse duration. Reducing the peak field to $E_0 = 1\,\rm{GV/m}$ relaxes the short pulse requirement to $\tau \sim 10\,\rm{ps}$ at the cost of weaker focusing and smaller $K_{\rm{und}}$. Even though the damage threshold restricts the drive-laser power, pulse front tilted laser beams~\cite{sas_ref.Wei.2017} offer an elegant solution to maximize the interaction length without increasing the pulse duration $\tau$. In exactly the same way tilted pulse front lasers enhance the energy gain in accelerating grating structures~\cite{sas_ref.Cesar.2018} the technique can be applied for DLA undulators as well.
    
    For both designs the performance critically depends on the maximum achievable focusing constant $K\left(\alpha\right)$. The ARES electron beam specified with $\varepsilon_{\rm{y}} = 1\,\rm{nm}$ geometric emittance reaches for $E_0 \sim 10\,\rm{GV/m}$ a beam size of $\sigma_{\rm{y}} \approx 1\,\rm{\upmu m}$ inside the DLA structure. Thus, non-negligible particles losses in the dielectric material of the beam channel with $\Delta y = 1.2\,\rm{\upmu m}$ gap width will occur. At $E_0 \sim 1\,\rm{GV/m}$ the expected beam size is about twice as large. Consequently, maximizing the laser field strength $E_0$ is a crucial aspect for the undulator's efficacy. The achievable photon wavelength covers a spectral range from XUV to soft X-ray radiation. For $E_0 \sim 10\,\rm{GV/m}$ the DLA undulator light source reaches up to $\hat{P}_{\rm{rad}} \sim 100 \,\rm{W}$ peak power, converting to $E_{\rm{rad}} \sim 0.1\,\rm{pJ}$ photon pulse energy per $0.5$~pC, $0.75\,\rm{fs}$ electron bunch. 
    \begin{table}
        \centering
            	\begin{tabular*}{\linewidth}{l@{\extracolsep{\fill}} r r r r}
		\hline
		 Laser Field Strength $E_{0}$ 	& \multicolumn{2}{c}{$10\,\rm{GV/m}$} 	& \multicolumn{2}{c}{$1\,\rm{GV /m}$} \\ 
		 Operation Mode 	& async. 	& sync. 	& async. 	& sync. \\ 
		\hline
		Und. Wavelength $\lambda_{\rm{und}}$ / $\rm{mm}$	&$0.55$	&$0.55$	&$1.00$	&$1.00$\\
		Und. Parameter $K_{\rm{und}}$	&$0.30$	&$0.30$	&$0.06$	&$0.06$\\
		Rad. Wavelength $\lambda_{\rm{rad}}$ / $\rm{nm}$	&$6.49$	&$6.49$	&$11.35$	&$11.35$\\
		Rad. Peak Power $\hat{P}_{\rm{rad}}$ / $\rm{W}$	&$140.1$	&$140.1$	&$2.9$	&$2.9$\\
		Opt. Betafunction $\hat{\beta}_{\rm{max}}$ / $\rm{mm}$	&$0.95$	&$1.07$	&$3.89$	&$4.75$\\
		Beam Size (rms) $\sigma_{\rm{y}}$ / $\rm{\mu m}$	&$0.98$	&$1.04$	&$1.97$	&$2.18$\\
		Synchronous Phase $\varphi_{\rm{s}}$ / $\rm{deg}$	&-	&$38.2$	&-	&$38.2$\\
		\hline
	\end{tabular*}

        \caption{\label{tab:design_undulator_experiment}Specifications of both DLA undulator designs for the DESY-ARES accelerator. The damage threshold of the silica grating limits the drive-laser amplitude to $E_0 = 10\,\rm{GV/m}$. The values at $E_0 = 1\,\rm{GV/m}$ serve for comparison.}
    \end{table}

    \section{\label{sec:particle_tracking}Particle Tracking Simulations}
    Limited by the laser damage threshold of silica, we assume $10\,\rm{GV/m}$ field strength for a sufficiently short, pulse front tilted drive-laser illumination. With a total length of $z = 1.25\,\rm{cm}$, the dimensions of the simulated DLA undulator lattice are similar to the typical size of commercially available transmission spectrometer gratings~\cite{sas_ref.IbsenPhotonics.2020}. In order to determine the Courant-Snyder parameters $\bm{\eta}$ of the injected electron distribution, the numerical studies initially consider a Gaussian phase space distribution with an idealized, hypothetical emittance of $\varepsilon_{\rm{y}} = 25\,\rm{pm}$ and $\varepsilon_{\rm{x}} = \varepsilon_{\rm{z}} = 0$ represented by $N_p = 10^5$ macroparticles. Based on the matched parameters $\bm{\eta}$, a subsequent, fully numeric simulation study provides the beam dynamics of the experimental setup for the ARES electron beam with $\varepsilon_{\rm{x}} = \varepsilon_{\rm{y}} = 1\,\rm{nm}$ transverse emittance and $\sigma_{\rm{t}} = 0.75\,\rm{fs}$ bunch length.
    
\begin{figure}
        \includegraphics{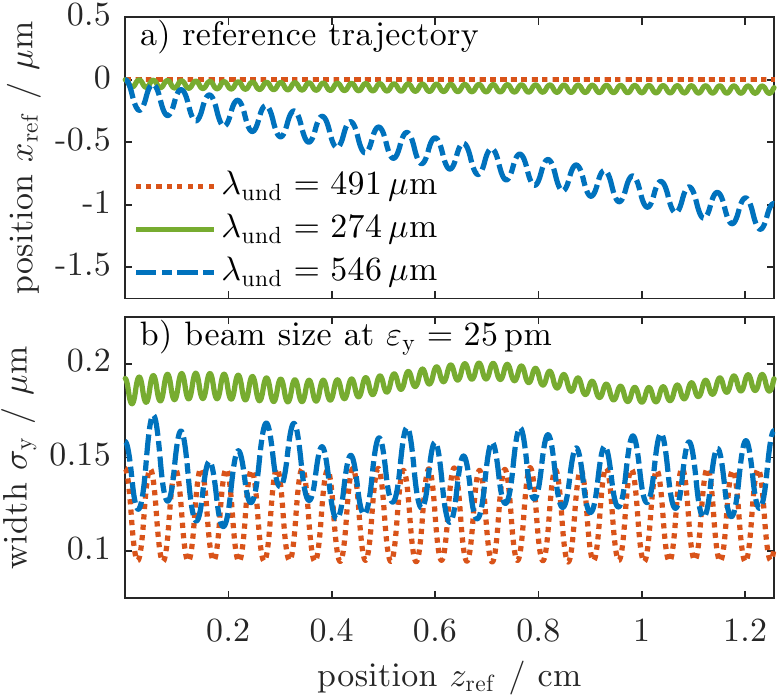}
        \caption{\label{figure10_schmid_prab} DLAtrack6D simulations of the reference trajectory in a) and transverse rms. size in b) of an electron beam with $\varepsilon_{\rm{y}} = 25\,\rm{pm}$ and $\varepsilon_{\rm{x}} = \varepsilon_{\rm{z}} = 0 $ emittance for three different setups of the synchronous DLA undulator lattice. The setup for $\varphi_s = 90^{\circ}$ synchronous phase with $\lambda_{\rm{und}} = 491\,\rm{\upmu m}$ provides maximum focusing but no transverse deflection. The other setups compare the beam dynamics for $\varphi_s = 38.2^{\circ}$ in two lattices designs with different undulator wavelengths $\lambda_{\rm{und}}$.}
    \end{figure}

    First, we employ the semi-analytical particle tracking code \emph{DLAtrack6D}~\cite{sas_ref.Niedermayer.2017} which provides a fast opportunity to investigate and optimize the non-linear beam dynamics inside the DLA undulator lattice. Figure~\ref{figure10_schmid_prab}\,a) compares the reference trajectory for three different setups of a synchronous DLA undulator lattice as shown in Fig.~\ref{figure3_schmid_prab}. According to \eqref{eqn:undulator_parameter_sync} and \eqref{eq:FODO:Hillsequation}, the transverse deflection $K_{\rm{und}}$ and the focusing strength $K_{\rm{\alpha}}$ of the undulator lattice depend on the synchronous phase $\varphi_s$ and the undulator period $\lambda_{\rm{und}}$.
    Figure.~\ref{figure10_schmid_prab}\,b) shows the corresponding beam size $\sigma_y$ for the reference electron distribution with $\varepsilon_{\rm{y}} = 25\,\rm{pm}$ emittance. The particle trajectory for $\lambda_{\rm{und}} = 491\,\rm{\upmu m}$ and $\varphi_s = 90^{\circ}$ (orange) corresponds to a purely focusing setup without transverse deflection. As visible in the strictly periodic $\sigma_y$-envelope function, perfectly matched injection was attained.
    
    The simulation results for $\lambda_{\rm{und}} = 546\,\rm{\upmu m}$ and $\lambda_{\rm{und}} = 274\,\rm{\upmu m}$ demonstrate that for the synchronous phase $\varphi_s = 38.2^{\circ}$ (cf. Tab.~\ref{tab:design_undulator_experiment}) the DLA lattice provides both deflection in Fig.~\ref{figure10_schmid_prab}\,a) and periodic focusing. According to \eqref{eqn:synchronous_phase_referenceparticle}, the transverse deflection in $x$-direction affects $\varphi_{\rm{s}}$ which, if not compensated, results in a drifting synchronous phase. The investigation of the dynamics in tilted gratings in Ref.~\cite{sas_ref.Niedermayer.2017} shows that a deflection induced phase drift converts to a coherent oscillation in the $x$-direction as well. However, the longitudinal period $\lambda_{\rm{u}}$ in Ref.~\cite{sas_ref.Niedermayer.2017} is larger than the undulator period $\lambda_{\rm{und}}$. Appropriate adjustment of the drift spaces allows to mitigate the phase drift which the transverse deflection along the $x$-axis induces in \eqref{eqn:synchronous_phase_referenceparticle}. A numerical optimization of the DLA lattice design yields a consecutive alternation of the drift space lengths by $\delta \lambda_{\rm{z}} \approx +9\%/-7\%$ for $\lambda_{\rm{und}} = 546\,\rm{\upmu m}$ and $\delta \lambda_{\rm{z}} \approx +2\%/-2\%$ for $\lambda_{\rm{und}} = 274\,\rm{\upmu m}$. Nevertheless, a constant drift motion in $x$-direction, as visible for the larger undulator period $\lambda_{\rm{und}} = 546\,\rm{\upmu m}$, remains. Moreover, slight mismatch caused by the non-linear field distribution of the DLA structure overlay the strictly periodic envelope oscillation of the beam size $\sigma_y$ with periodicity $\lambda_{\rm{und}}$ in Fig.~\ref{figure10_schmid_prab}\,b). Since the FODO segment length corresponding to $\lambda_{\rm{und}} = 546\,\rm{\upmu m}$ was obtained from an optimization of the Courant-Snyder parameters for the given DLA cell in Fig.~\ref{figure2_schmid_prab} with $\alpha=27^\circ$, deviating from that segment length to $\lambda_{\rm{und}} = 274\,\rm{\upmu m}$ results in a sub-optimal (larger) beam size.
    
\begin{figure}
        \includegraphics{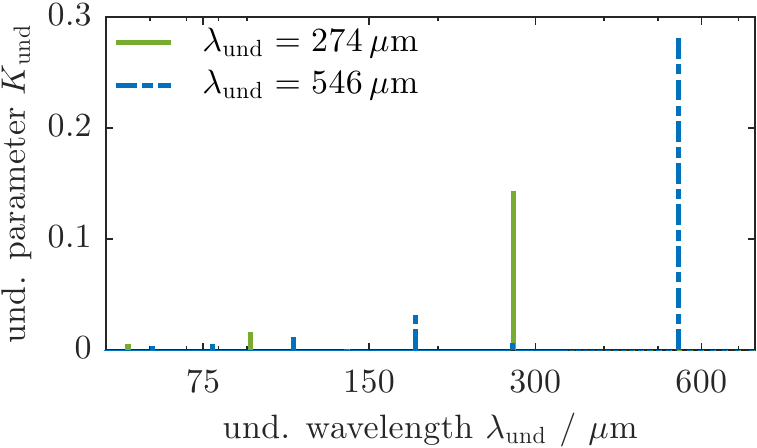}
        \caption{\label{figure11_schmid_prab}
        Fourier analysis of the reference trajectory $x^{\prime}\left(z\right)$ in order to determine the contributing oscillation wavelengths $\lambda_{\rm{und}}$ and the effective undulator parameter $K_{\rm{und}}$ for the synchronous DLA undulator lattices.}
    \end{figure}

    A Fourier analysis of the reference trajectory yields the spectral components of the transverse particle deflection $x^{\prime}\left(z\right)$. Figure~\ref{figure11_schmid_prab} shows the numerical results for the effective undulator parameter $K_{\rm{und}}$ of the contributing oscillation wavelengths $\lambda_{\rm{und}}$. In very good agreement with \eqref{eqn:undulator_parameter_sync}, the fundamental harmonic oscillation in the DLA lattice design for $\lambda_{\rm{und}} = 546\,\rm{\upmu m}$ corresponds to an effective undulator parameter of $K_{\rm{und}} \approx 0.29$. Furthermore, at $\lambda_{\rm{und}} \in \left\lbrace 182\,\rm{nm}, 109\,\rm{nm},\dots \right\rbrace$ the spectrum shows contributions of uneven higher harmonics which are typical for the synchronous undulator setup. According to \eqref{eqn:undulator_parameter_sync}, the undulator parameter scales linearly with the undulator wavelength such that $\lambda_{\rm{und}} = 274\,\rm{\upmu m}$ results in $K_{\rm{und}} \approx 0.14$.
    
\begin{figure}
        \includegraphics{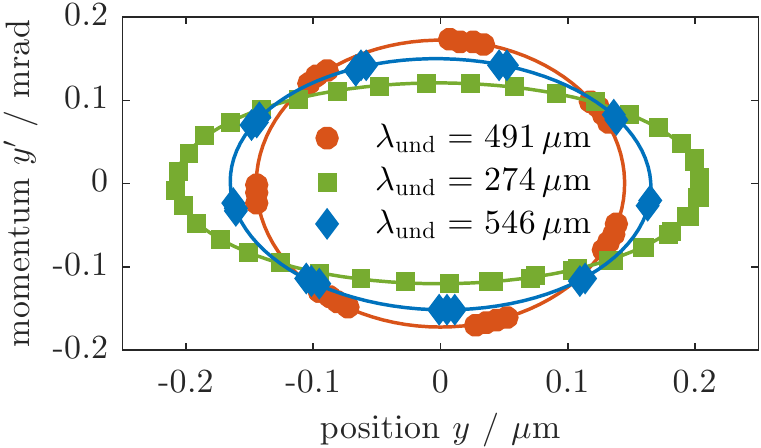}
        \caption{\label{figure12_schmid_prab} Poincar\'{e} plots for the three synchronous DLA undulator lattices based on the phase space coordinates $\left(y,y^{\prime}\right)$ of particles with $\varepsilon_{\rm{y}} = 25\,\rm{pm}$ single-particle emittance evaluated after each FODO period $\lambda_{\rm{und}}$}
    \end{figure}

    Figure \ref{figure12_schmid_prab} shows that the phase space ellipses of the three different setups can be obtained from Poincar\'{e} cross-sections (after each FODO period $\lambda_{\rm{und}}$) of the phase space coordinates $\left(y,y^{\prime}\right)$ for a particle with $\varepsilon_{\rm{y}} = 25\,\rm{pm}$ single-particle emittance. The solid lines are obtained from fitting the points to an ellipse, which shows that the single particle emittance as a Courant-Snyder invariant is preserved. Moreover, matching at injection is at least approximately confirmed.
    As expected by design, the purely focusing DLA lattice with $\lambda_{\rm{und}} = 491\,\rm{\upmu m}$ and $\varphi_s = 90^{\circ}$ provides the smallest transverse beam size. Furthermore, the numerically computed maxima of the betatron function $\hat{\beta}_{\rm{max}} = 1.1\,\rm{mm}$ for $\lambda_{\rm{und}} = 546\,\rm{\upmu m}$ and $\hat{\beta}_{\rm{max}} = 1.7\,\rm{mm}$ for $\lambda_{\rm{und}} = 274\,\rm{\upmu m}$ fit very well to the predictions of the linearized focusing model $\eqref{eq:FODO:Hillsequation}$.
    
\begin{figure}
        \includegraphics{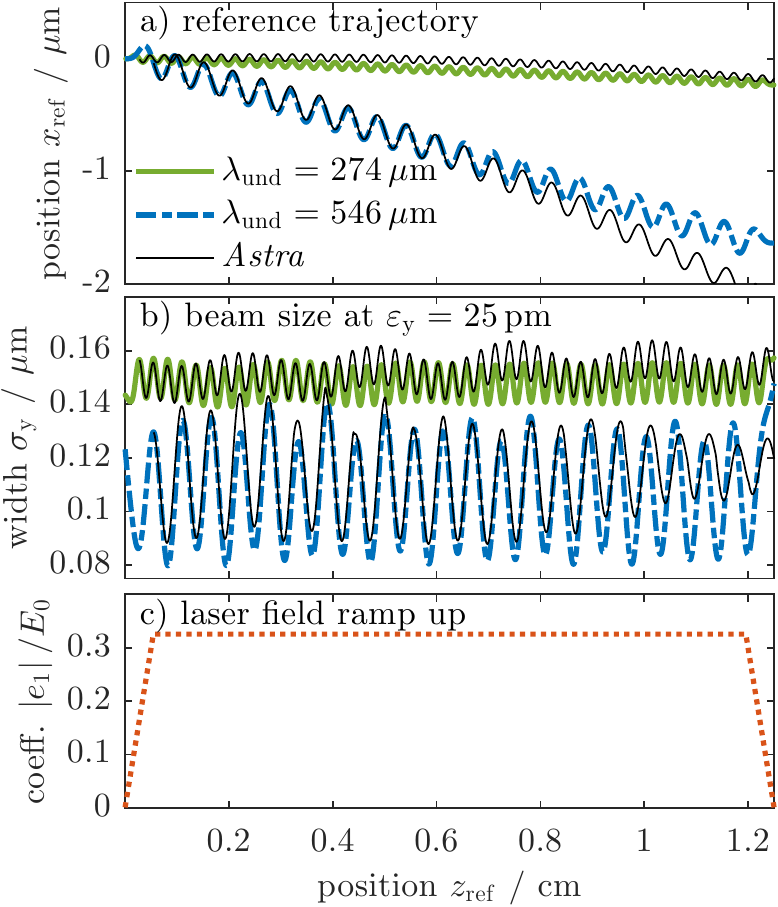}
        \caption{\label{figure13_schmid_prab} \emph{DLAtrack6D} simulations of the reference trajectory in a) and transverse rms. size in b) of an electron beam with $\varepsilon_{\rm{y}} = 25\,\rm{pm}$ and $\varepsilon_{\rm{x}} = \varepsilon_{\rm{z}} = 0 $ emittance for two operation modes of the asynchronous DLA undulator with linear laser field ramp-up shown in c). The  \emph{Astra}~\cite{sas_ref.KlausFloettmann.2017} simulations show the results of a fully numeric, time-dependent beam dynamics model calculated with particle tracking in a 3D field map of the DLA cell from \emph{CST Studio}~\cite{sas_ref.3DS.2021}. }
    \end{figure}

    Figure~\ref{figure13_schmid_prab} shows \emph{DLAtrack6D} simulations of the asynchronous DLA undulator lattice. According to \eqref{eqn:hamilton_vectorpotenital_transversex}, electrons injected into the asynchronous DLA undulator at the laser phase of maximum focusing, $\varphi_{\rm{s}} = 90^{\circ}$, acquire a transverse momentum offset in $x$-direction. However, simulations show that a linear ramp-up of the laser field amplitude across one undulator period $\lambda_{\rm{und}}$ as shown in Fig.~\ref{figure13_schmid_prab}\,c) allows to mitigate the resulting drift motion. The intensity distribution of the drive-laser pulse in a practical DLA experiment automatically introduces a reduction of field strength at the entrance and exit of the DLA undulator. In the numerical model, we simplified the ramp-up as linear. However, the model can easily be extended to a more realistic, e.g. Gaussian, shape. Regarding the deflection and focusing properties, the simulation results are to be interpreted in the same way as the aforementioned synchronous DLA setup. Evaluation of the numerical simulation data for $\lambda_{\rm{und}} = 546\,\rm{\upmu m}$ yields $K_{\rm{und}} \approx 0.28$ and $\hat{\beta}_{\rm{max}} \approx 1\,\rm{mm}$ and, thereby, reproduces the deflection and focusing properties listed in Tab.~\ref{tab:design_undulator_experiment}.
    
    In order to validate the results of the semi-analytic \emph{DLAtrack6D} model for the asynchronous setup, Fig.~\ref{figure13_schmid_prab} additionally shows simulation results of the fully numeric particle tracking code \emph{Astra}~\cite{sas_ref.KlausFloettmann.2017}. The \emph{Astra} simulation uses a time-dependent representation of the laser field distribution inside the beam channel of the DLA lattice (see Appx.~\ref{apx:astra_3dfieldmaps} describing the concatenation of single cell fields for the entire structure). Since modeling the ramp-up of the field strength would extend the computational domain to the memory limit of \emph{Astra}, the tracking simulation starts in the fully periodic region of the DLA lattice for $z > \lambda_{\rm{und}}$. The initial particle distribution is obtained from a DLAtrack6D simulation of the ramp-up section. The detuning required for the asynchronous undulator was chosen according to \eqref{eqn:hamilton_vectorpotenital_undulatorwavelength} as $\delta \lambda_0 \approx 7.3\,\rm{nm}$ for $\lambda_{\rm{und}} = 546\,\rm{\upmu m}$ and $\delta \lambda_0 \approx 14.5\,\rm{nm}$ for $\lambda_{\rm{und}} = 274\,\rm{\upmu m}$. The electromagnetic field model uses the same grating structure in both cases and the frequencies $f_0 = 150.45\,\rm{THz}$ and $f_0 = 150.99\,\rm{THz}$, respectively. Comparing the \emph{Astra} and \emph{DLAtrack6D} simulation approaches in Fig.~\ref{figure13_schmid_prab} shows that the undulator wavelength, the oscillation amplitude, and the beam width agree very well. Hence, we consider designing both synchronous and asynchronous lattices with \emph{DLAtrack6D} as legitimate, while a cross-check with a much slower full field simulation tool should be reserved for a finalized design.
    
    With the current ARES beam parameters, losses in the beam transport through the investigated DLA undulator lattice are unavoidable. Limited by the focusing strength of the DLA lattice in $y$ direction, roughly $50\%$ of the initial particle bunch are lost during injection. Because the spatial extend of the beam is not negligibly small compared to the characteristic dimensions $2\pi/k_{\rm{x}}$ and $2\pi/k_{\rm{z}}$ of the DLA cells, it is impracticable to match the entire electron bunch to the optics functions of the DLA FODO lattice and there is a dynamical aperture that is smaller than the physical aperture. 

\begin{figure}
        \includegraphics{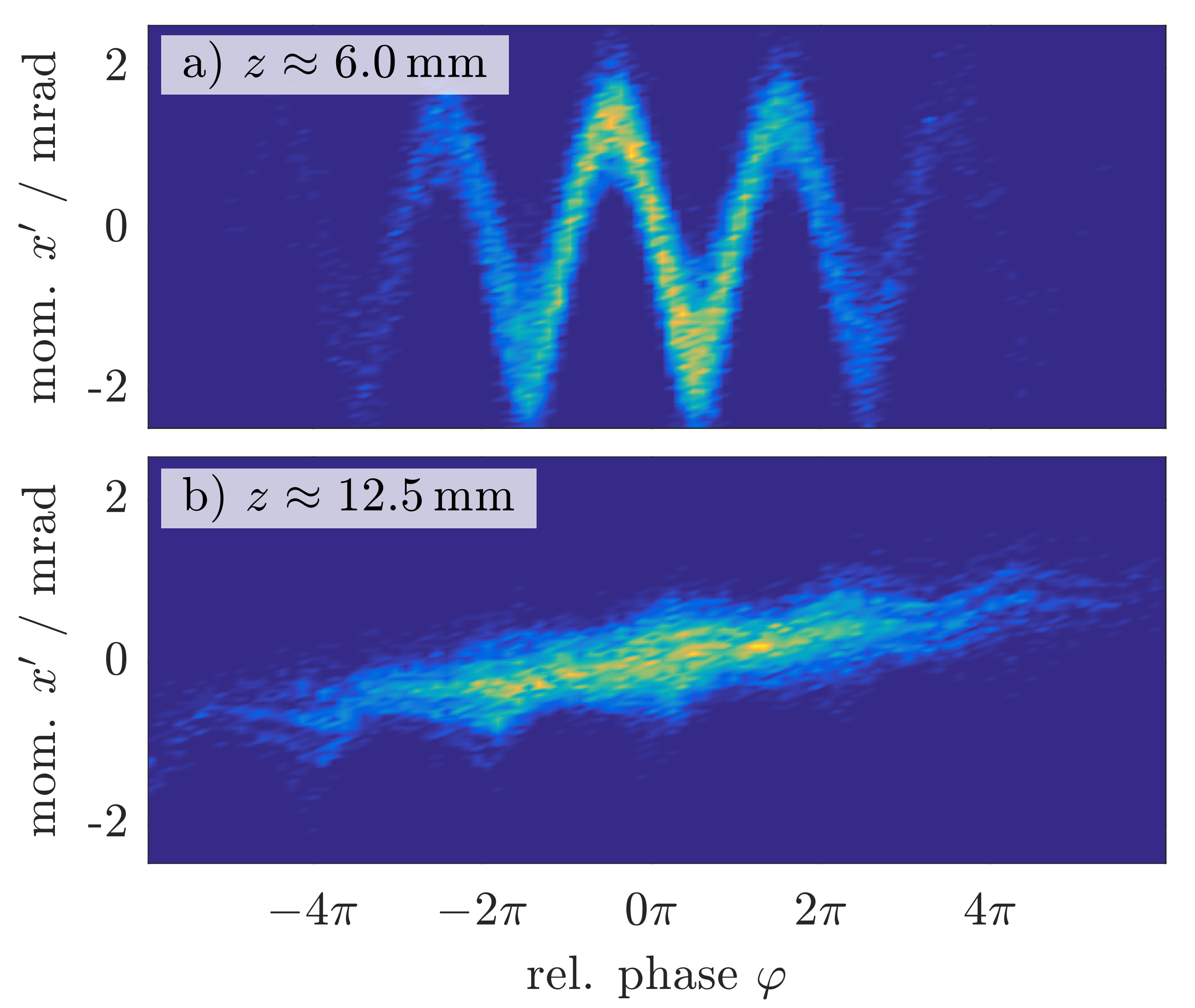}
        \caption{\label{figure14_schmid_prab} \emph{DLAtrack6D} simulation of the $\lambda_{\rm{und}} = 546\,\rm{\upmu m}$ asynchronous DLA undulator with realistic ARES injection parameters. The plots show the transverse momentum vs. the relative phase (cf. \eqref{eqn:asynchronous_phase}) at two different $z$-positions.}
    \end{figure}

    Figure~\ref{figure14_schmid_prab} shows the transverse momentum $x^{\prime}$ vs. $\varphi$ (cf. \eqref{eqn:asynchronous_phase}) distribution in the center and the end of the DLA undulator. Since the dynamical evolution of each particle described by \eqref{eqn:hamilton_vectorpotenital_transversex} depends crucially on the initial parameters $x_0$ and $c t_0$, the oscillations of a beam that is not microbunched are incoherent and the generated photons subsequently are also incoherent. In order to obtain coherence, microbunching prior to injection is required.
    
    The electron distribution at the rear end of the asynchronous DLA undulator in Fig.~\ref{figure14_schmid_prab}\,b) provides approximately $17\%$ transmission with the ARES beam injection. Additionally to the injection itself, most of the particle losses occur within the first couple undulator wavelengths $z \lesssim 2 \lambda_{\rm{und}}$. Doubling the total length in the \emph{DLAtrack6D} simulation to $L \approx 2.5\,\rm{cm}$ still provides $17\%$ and for $L \approx 6\,\rm{cm}$ approximately $16.5\%$ of the injected electrons remain. Hence, scalability of the confinement, within its limitations of the dynamic aperture, is confirmed. 
    
\begin{figure}
        \includegraphics{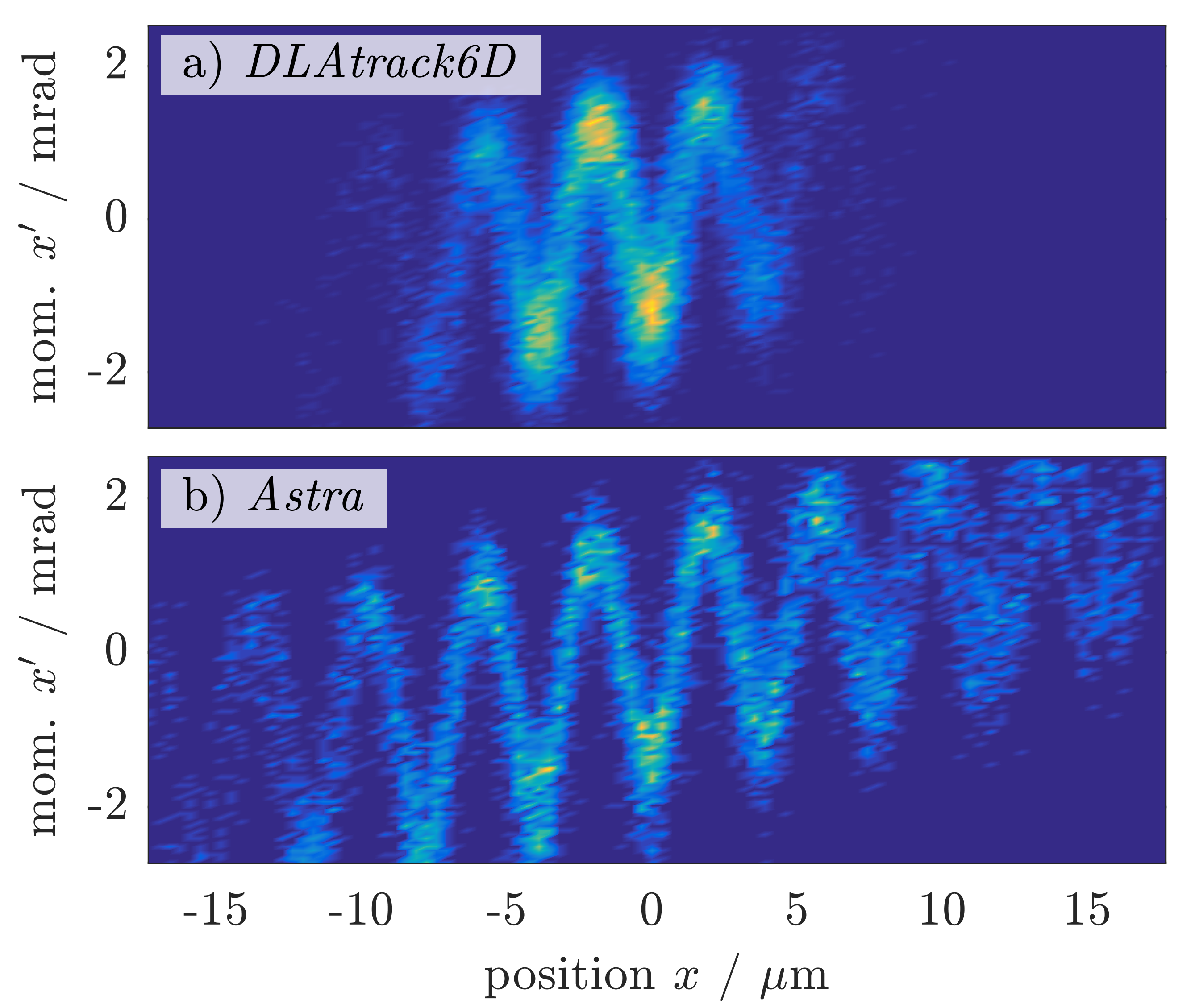}
        \caption{\label{figure15_schmid_prab} Comparison of horizontal phase space from \emph{DLAtrack6D} and \emph{Astra} at $z \approx 9.3\,\rm{mm}$ in the asynchronous DLA undulator lattice for $\lambda_{\rm{und}} = 546\,\rm{\upmu m}$.}
    \end{figure}

\begin{figure}
        \includegraphics{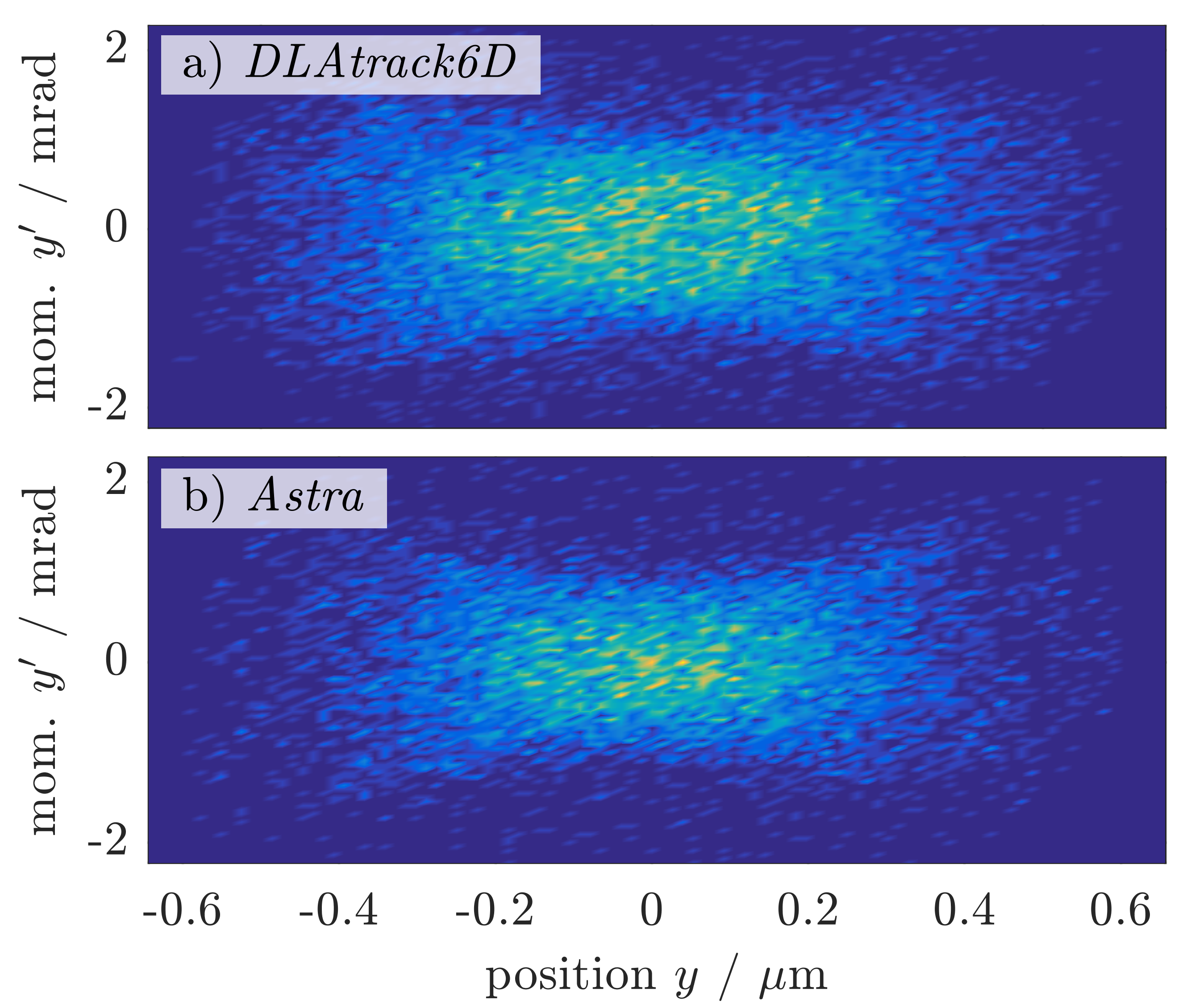}
        \caption{\label{figure16_schmid_prab} Comparison of vertical phase space from \emph{DLAtrack6D} and \emph{Astra} at $z \approx 9.3\,\rm{mm}$ in the asynchronous DLA undulator lattice for $\lambda_{\rm{und}} = 546\,\rm{\upmu m}$.}
    \end{figure}

    Figures~\ref{figure15_schmid_prab} and \ref{figure16_schmid_prab} compare a) \emph{DLAtrack6D} and b) \emph{Astra} simulation results for the transverse phase spaces $x-x^{\prime}$ and $y-y^{\prime}$ at $z \approx 9.3\,\rm{mm}$. The laser field interaction imprints the transverse grating periodicity with $\lambda_{x} = 2\,\rm{\upmu m} / \tan{27^{\circ}} \approx 3.9\,\rm{\upmu m}$ on the electron beam's $x-x^{\prime}$ phase space distribution in Fig.~\ref{figure15_schmid_prab}. For the transverse $y-y^{\prime}$ coordinates in Fig.~\ref{figure16_schmid_prab}, the focusing properties of the DLA lattice maintain a spatially confined beam.
    
    While the results with regard to phase space shape are very similar, the \emph{Astra} simulation reaches only $8.3\%$ transmission rate as compared to $17\%$ in \emph{DLAtrack6D}. This discrepancy is related to edge field effects at the transverse boundaries of the beam channel $\left|y\right| \rightarrow 0.6\,\rm{\upmu m}$ which are not considered in the \emph{DLAtrack6D} model (the \emph{DLAtrack6D} model assumes a $\cosh$-shaped transverse profile, which is inaccurate close to the channel boundary). Furthermore, the \emph{Astra} simulation in Fig.~\ref{figure15_schmid_prab}\,b) predicts a larger beam divergence in $x$-direction than \emph{DLAtrack6D} in Fig.~\ref{figure15_schmid_prab}\,a). This indicates that the investigated DLA lattice design cannot completely compensate the transverse particle drift motion \eqref{eqn:hamilton_vectorpotenital_transversex} which the laser field imposes on the particles during injection. Since a mismatched beam injection can induce both particle losses as well as transverse drift motion, a fully numeric tracking optimization of the lattice design and injection properties might mitigate the these discrepancies. In an experiment, the injection parameters are usually tweaked manually for best performance.
    
\section{Conclusion}
    In summary, this study considers two different concepts of laser driven undulator designs based on either synchronous or asynchronous tilted-grating APF-DLA lattices. The theoretical investigation of the beam dynamics in such structures contributes to a collection of basic design guidelines for experimental setups, which can be implemented for example at the DESY-ARES accelerator. First, a wave-optical simulation of the silica grating allows to optimize the DLA cell geometry for maximum interaction between the electron beam and the drive-laser field. Second, the analytical analysis of the beam dynamics yields approximate formulas for the undulator parameter and the focusing strength. Application of these formulas shows that the DLA undulator can be operated as an APF focusing channel. Third, the focusing properties of the DLA undulator impose direct restrictions on the achievable undulator wavelength. For the parameters considered in this study, the optimization of the beam focusing yields an undulator wavelength of approximately $\lambda_{\rm{und}} = 0.6\,\rm{mm}$ which will generate XUV and soft X-ray photons. Fourth, semi-analytical \emph{DLAtrack6D} simulations and fully numeric particle tracking e.g. with \emph{Astra} facilitate to match the Courant-Snyder parameters of the injected electron bunch to the beam optics functions of the DLA lattice. Beam dynamics simulations are a crucial step in the design process, since mismatched Courant-Snyder parameters would result in excessive particle losses. Last but not least, a comparison of both synchronous and asynchronous undulator concepts reveals the advantages and drawbacks of the two approaches. The asynchronous DLA undulator is very promising since its experimental setup is comparably simple and can be implemented by bonding two commercially available diffraction gratings together. Synchronous DLA undulators require a dedicated grating structure design and fabrication, but offer more design options to optimize the beam dynamics. In order to obtain coherent radiation output, the beam has to be bunched on the optical scale, such that it undulates coherently. Moreover, its transverse size (or better transverse emittance) needs to be sufficiently small to fit in one of the "transverse buckets". Such a matching is possible at ARES, when the accelerator reaches its full performance~\cite{sas_ref.Mayet.2018}. With the larger bunch as considered here, the electrons will not oscillate with a proper phase relation and therefore incoherent radiation is obtained. Detecting this radiation and characterizing its spectral properties is, however, already an experimental challenge worthy of addressing.
    
    Although the predicted power output is miniscule, the simple design of an asynchronous DLA undulator lattice qualifies it as a promising concept for a demonstration experiment at ARES. A setup based on two commercially available silica diffraction gratings like in Ref.~\cite{sas_ref.IbsenPhotonics.2020} allows to conduct a comparably simple radiation generation experiment. However, the size limitation of the beam channel and the laser induced beam divergence will result in significant electron losses inside and behind the DLA structure. Utilizing a $\lambda_0 = 5\,\rm{\upmu m}$ MIR drive-laser currently developed at DESY-ARES rather than the current $2\,\rm{\upmu m}$ one offers one possibility to lift design restrictions. Due to the scale in-variance of Maxwell's equations, the relative electron beam size in the gap of the DLA undulator decreases like $\sigma_y / \Delta y \propto {\lambda_0}^{-3/4}$ which reduces the particle losses in the structure. Likewise the radiation wavelength increases like $\lambda_{\rm{rad}} \propto {\lambda_0}^{1/2}$. However, such setups require different grating materials since silica looses its transparency for MIR wavelengths. Another approach for DLA experiments considers the injection of preconditioned electron beams in form of micro-bunch trains~\cite{sas_ref.Mayet.2018}. Utilizing tailored electron beams for DLA undulators is a promising concept, because both the particle transmission rate as well as the spectral coherence of the X-ray emission potentially profit. Considering the impact on large scale X-ray light sources, it is rather unlikely that tilted DLA gratings will become a replacement for conventional magnetic undulators. For instance, the E-XFEL utilizes a $Q = 1\,\rm{nC}$ bunch charge, $\varepsilon_{\rm{x}}^{\rm{n}} = 1.4\,\rm{mm\,mrad}$ normalized emittance electron beam in order to generate $\hat{P}_{\rm{rad}} \sim 100\,\rm{GW}$ peak radiation power with ultra large brilliance~\cite{sas_ref.Altarelli.2006}. Hence, the particle beam properties deviate significantly from the working range of the investigated DLA structures. However, small scale DLA setups open new research prospects by pushing the photon pulse duration from $\sigma_{\rm{t}} \sim 100\,\rm{fs}$~\cite{sas_ref.Altarelli.2006} into the sub-femtosecond regime. Especially the combination of DLA undulators with DLA electron accelerators is a promising concept, since a complete light source could be integrated on a single chip. Such a device offers the potential to enable typical large scale pump-probe experiments in table-top sized setups.
    
\begin{acknowledgments}
This work is funded by the German Federal Ministry of Education and Research (BMBF Grant No. 05K19RDE) and the Gordon and Betty Moore Foundation (Grant No. GMBF4744, ACHIP). We would like to thank the colleagues from DESY for providing the ARES beam parameters.
\end{acknowledgments}

\appendix
\section{\label{apx:asynchronous_vectorfield_ansatz}Electromagnetic Field Model for a Tilted DLA Grating}
    
    The following section contains supplementary information regarding an ansatz for the scaled vector potential in an asynchronous DLA undulator. The geometry and symmetry of the DLA grating structure motivate the analytical form of the vector potential ansatz \eqref{eqn:representation_vetorpotential} in Sec.~\ref{sec:undulator_lattice}. 
    
    Considering the special case of a non-tilted grating with $\alpha = 0$, the wave vector of the electromagnetic field distribution in the DLA beam channel is perpendicular to the $x$-direction $\bm{k} \perp \bm{\hat{x}}$. Since the DLA grating geometry remains invariant for a mirror reflection in the $y-z$-plane~\cite{sas_ref.Naranjo.2012}, the electromagnetic field decomposes into TE $\left(E_{\rm{y}},E_{\rm{z}},B_{\rm{x}}\right)$ and TM $\left(B_{\rm{y}},B_{\rm{z}},E_{\rm{x}}\right)$ polarized modes~\cite[pp.~37-39]{sas_ref.Joannopoulos.2008} which each fulfill Maxwell's equations. Hence, a reasonable ansatz for a fundamental spatial harmonic of the vector potential $\bm{A}_{\alpha}\left(x,y,z,ct\right)$ in a non-tilted DLA structure with grating wave-number $k_{\rm{g}}$ and $\alpha = 0$ is 
    \begin{equation}
        \bm{A}_0\left(x,y,z,ct\right) = \begin{pmatrix} A_{\rm{TM}} &\cosh{\left(k_{\rm{y}} y\right)} \sin{\left(k_{\rm{g}} z - k_{\rm{0}} ct\right)}\\ -\frac{k_{\rm{g}}}{k_{\rm{y}}} A_{\rm{TE}} &\sinh{\left(k_{\rm{y}} y\right)} \cos{\left(k_{\rm{g}} z - k_{\rm{0}} ct\right)}\\ A_{\rm{TE}} &\cosh{\left(k_{\rm{y}} y\right)} \sin{\left(k_{\rm{g}} z - k_{\rm{0}} ct\right)} \end{pmatrix}
    \end{equation}
    with the corresponding mode amplitudes $A_{\rm{TE}}$ and $A_{\rm{TM}}$. Note that the ansatz represents a $y$-symmetric field distribution which is periodic in longitudinal $z$-direction and decays exponentially towards the center of the beam channel at $y = 0$.
    
    A rotation $\hat{R}_{\rm{\hat{y}}}\left(\alpha\right)$ around the $y$-axis transforms the vector potential $\bm{A}_0\left(x,y,z,ct\right)$ into the coordinate system $\left(\tilde{x},\tilde{y},\tilde{z},ct\right)$ of an electron beam propagating at a tilt angle $\alpha$ with respect to the $z$-axis of the non-tilted grating. For the corresponding mode amplitudes in the new coordinates the rotation $\hat{R}_{\rm{\hat{y}}}\left(\alpha\right)$ yields
    \begin{equation}
        \begin{pmatrix} A_{\rm{\tilde{x}}} \\ A_{\rm{\tilde{y}}} \\ A_{\rm{\tilde{z}}} \end{pmatrix} = \begin{pmatrix} A_{\rm{TM}} \cos{\alpha} + A_{\rm{TE}} \sin{\alpha} \\ -\frac{k_{\rm{g}}}{k_{\rm{y}}} A_{\rm{TE}} \\ A_{\rm{TE}} \cos{\alpha} - A_{\rm{TM}} \sin{\alpha}\end{pmatrix}
    \end{equation}
    and, thereby, defines the polarization vector $\bm{\xi}$ in \eqref{eqn:representation_vetorpotential}. Note that the components of $\bm{\xi}$ are not fully independent since $\bm{A}_{\alpha}$ is a superposition of the TE and TM modes of the dielectric grating. Applying the identity $k_{\rm{\tilde{z}}} = k_{\rm{g}} \cos{\alpha}$, the transformation of the spatial coordinates reproduces the relation between the transverse and longitudinal grating wave-numbers $k_{\rm{\tilde{x}}} = k_{\rm{\tilde{z}}} \tan{\alpha}$ from Ref.~\cite{sas_ref.Niedermayer.2017}. Finally, conversion of the laser field strength $E_0$ to the corresponding vector potential amplitude $\frac{E_0}{k_0 c}$ and scaling $\bm{A}_{\alpha}$ by $\frac{q}{m c}$ yields the scaled vector potential ansatz \eqref{eqn:representation_vetorpotential}.

\section{\label{apx:asynchronous_perturbation_theory}Perturbation Theory Applied to the Equations of Motion in an Asynchronous DLA}
    This section addresses a perturbation theory model for the beam dynamics in an asynchronous DLA undulator. In order to provide a reasonable approximation, the analytical model requires a relativistic electron beam and a reasonably small vector potential amplitude $a_0 \rightarrow 0$. In that case, the perturbation approach results in the approximation for the particle trajectories discussed in Sec.~\ref{sec:undulator_lattice}.
    
    Using the ansatz \eqref{eqn:representation_vetorpotential} in \eqref{eqn:hamilton_vectorpotenital_transversex}, Hamilton's equations
    \begin{equation}\label{eqn:appendix_hamilton_ODE}
        \frac{d}{dz} \begin{pmatrix} x \\ y \\ ct \\ p_{\rm{x}} \\ p_{\rm{y}} \\ -\gamma \end{pmatrix} = \begin{pmatrix} {\partial H}/{\partial p_{\rm{x}}} \\ {\partial H}/{\partial p_{\rm{y}}} \\ {\partial H}/{\partial \gamma} \\ -{\partial H}/{\partial x} \\ -{\partial H}/{\partial y} \\ -{\partial H}/{\partial ct} \end{pmatrix}
    \end{equation}
    for the canonical coordinates $x\left(z\right)$, $y\left(z\right)$, $ct\left(z\right)$, $p_{\rm{x}}\left(z\right)$, $p_{\rm{y}}\left(z\right)$, and $-\gamma\left(z\right)$ as functions of the independent variable $z$ yield a system of six coupled nonlinear first order ordinary differential equations (ODEs). Considering a relativistic electron beam with $\gamma \gg 1$, a series approximation for the right hand side of \eqref{eqn:appendix_hamilton_ODE} including terms $\mathcal{O}\left(1/\gamma^2\right)$ allows to simplify the ODEs to
    \begin{equation}\label{eqn:appendix_hamilton_ODE_gamma}
        \frac{d}{d z} \begin{pmatrix} x \\ \vdots \end{pmatrix} \approx \begin{pmatrix} \frac{p_{\rm{x}}- \xi_{\rm{x}} a_0 \cosh\left(k_{\rm{y}} y\right) \sin\left(k_{\rm{x}} x + k_{\rm{z}} z - k_{\rm{0}} ct\right)}{\gamma} \\ \vdots \end{pmatrix}\rm{.}
    \end{equation}
    
    For the DLA undulator discussed in Sec.~\ref{sec:undulator_lattice} the amplitude \eqref{eqn:hamilton_vectorpotential_amplitude} of the scaled vector potential is small $a_0 \ll 1$. Hence, the solutions of \eqref{eqn:appendix_hamilton_ODE_gamma} can be approximated considering the laser field interaction as a perturbation of the electron trajectory. If the laser is switched off, $a_0 = 0$, the unperturbed electron follows the trajectory of a free particle $x_0\left(z\right)$. In order to approximate the solution for $a_0 > 0$, the straightforward perturbation method~\cite[pp. 355-359]{sas_ref.Jazar.2021} substitutes
    \begin{align}
        a_0 &\rightarrow \epsilon \\
        x\left(z\right) &\rightarrow x_0\left(z\right) + \epsilon\,x_1\left(z\right) + \epsilon^2\,x_2\left(z\right) \\
        y\left(z\right) &\rightarrow y_0\left(z\right) + \epsilon\,y_1\left(z\right) + \epsilon^2\,y_2\left(z\right) \\
        &\vdots
    \end{align}
    in \eqref{eqn:appendix_hamilton_ODE_gamma} and sorts the resulting terms by increasing powers in $\epsilon$. Incremental computation of the perturbation terms $x_1\left(z\right)$, $x_2\left(z\right)$, etc. by reinserting the results of each previous calculation step in \eqref{eqn:appendix_hamilton_ODE_gamma} provides the beam dynamics corrections for laser field interaction. The procedure has been implemented in \emph{Mathematica}~\cite{sas_ref.WolframResearchInc..2021} which yields the approximation \eqref{eqn:hamilton_vectorpotenital_transversex} if all terms up to $\mathcal{O}\left(\epsilon^2\right)$ are taken into account.
    
    In most accelerator designs the relative energy gain of the particles $\frac{q E_0}{m c^2}$ per accelerating field wavelength $2\pi/k_0$ is comparably small. Thus, the fundamental assumption of the perturbation approach $a_0 \ll 1$ does not specifically apply to DLA undulators, but is also valid in the more general context of electromagnetic undulators like e.g. radio frequency wigglers as discussed in Ref.~\cite{sas_ref.Tran.1987}. However, the approximation \eqref{eqn:hamilton_vectorpotenital_transversex} fails describing the beam dynamics of a synchronous DLA undulator. If the synchronicity condition \eqref{eqn:phase_synchronous_condition} is fulfilled, the undulator wave number \eqref{eqn:hamilton_vectorpotenital_undulatorwavelength} vanishes and, therefore, the oscillation amplitude in \eqref{eqn:hamilton_vectorpotenital_transversex} diverges. The perturbation approach assuming small corrections $x_1\left(z\right)$, etc. is not valid anymore. However, the semi-analytical tracking approach for synchronous DLA structures in \emph{DLAtrack6D}~\cite{sas_ref.Niedermayer.2017} provides a suitable workaround to compensate the weakness of the perturbation theory model for $k_{\rm{und}} \rightarrow 0$. 
    
\section{\label{apx:astra_3dfieldmaps}Construction of 3D Field Maps for the \emph{Astra} Simulations}

    The fully numeric particle tracking simulation \emph{Astra}~\cite{sas_ref.KlausFloettmann.2017} uses 3D field maps of the drive-laser distribution in the beam channel of the DLA undulator lattice. An electromagnetic field simulation in \emph{CST Studio}~\cite{sas_ref.3DS.2021} provides the complex field amplitudes for the time-harmonic laser field on a hexahedral mesh in the computational domain. The numerical model applies a magnetic boundary condition $B_{\rm{x}} = B_{\rm{z}} = 0$ in the symmetry plane of the DLA cell at $y = 0$ in order to reduce the computational effort and ensure a mirror-symmetric field distribution.
    \begin{figure}
        \includegraphics{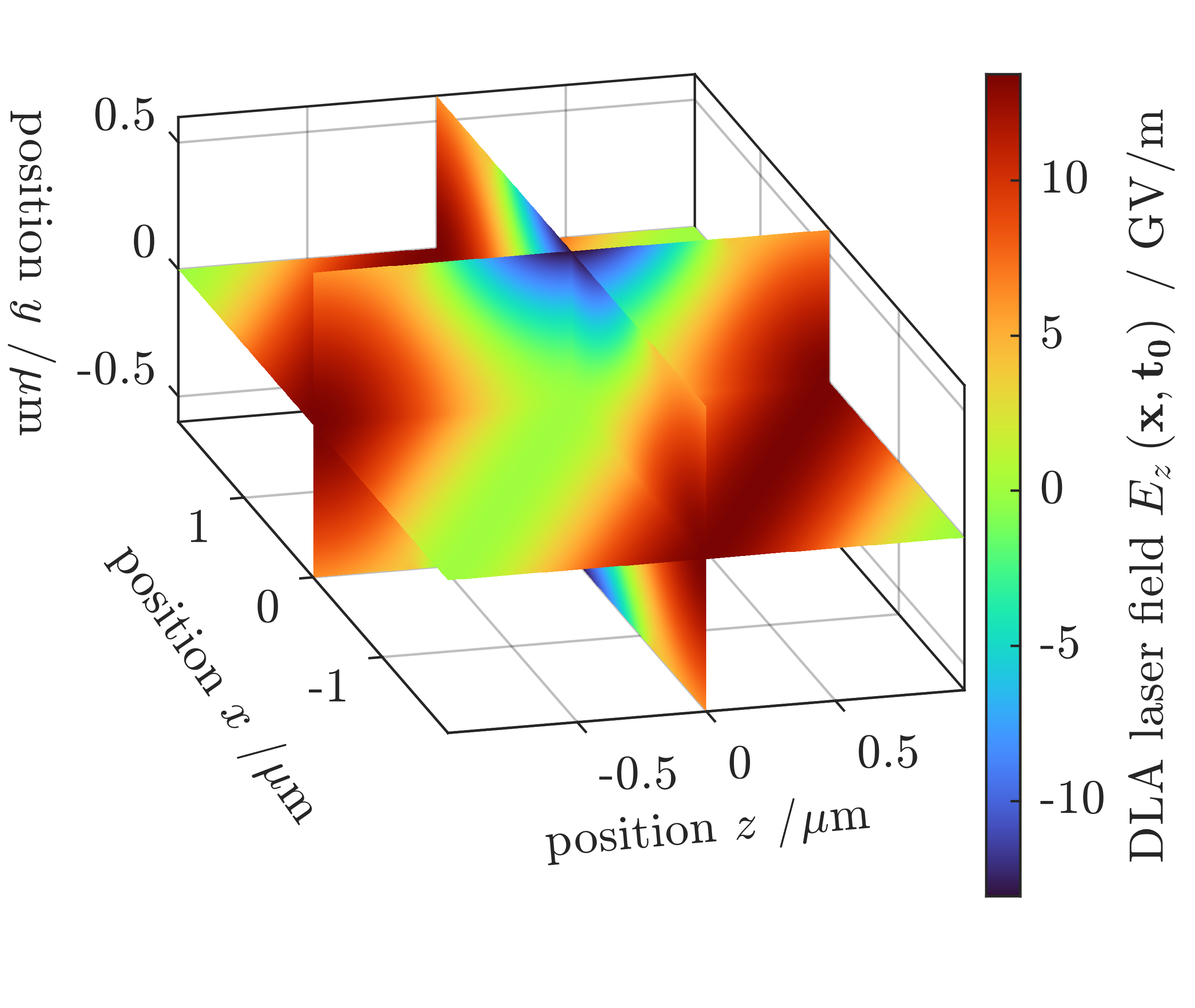}
        \caption{\label{figure17_schmid_prab} 3D field map of the laser distribution in the beam channel of a tilted silica DLA grating unit cell. The simulations conducted with the particle tracking code \emph{Astra}~\cite{sas_ref.KlausFloettmann.2017} use a periodic extension in order to model the field distribution of the full DLA lattice.}
    \end{figure}
    
    A post-processing step in \emph{Matlab}~\cite{sas_ref.TheMathworksInc..2016} transforms the output of \emph{CST Studio} to the input file format for 3D cavity field maps in \emph{Astra}. First, mirroring the numerical simulation results at the $y=0$ boundary extends field map to the upper half-space and provides the unit cell of the DLA grating. Figure~\ref{figure17_schmid_prab} shows the electromagnetic field in one unit cell of a tilted silica grating. Second, a periodic repetition in $x$-direction extends the DLA unit cell to several transverse periods of the DLA lattice. This step is cumbersome but necessary, because \emph{Astra} does not provide a periodic repetition of cavity field maps in transverse direction. However, periodic repetition in longitudinal $z$-direction is possible, such that \emph{Astra} can model the complete asynchronous DLA undulator with upwards of $6000$ silica grating periods in a single tracking simulation. The laser distribution of the DLA undulator lattice enters the \emph{Astra} tracking simulation in form of time-dependent accelerator cavity field maps.

\bibliography{DesignStudyDielectricLaserUndulator_SASchmid_UNiedermayer.bib}

\end{document}